\documentclass[preprint,12pt,3p,times]{elsarticle}
\usepackage{setspace}
\doublespace

\usepackage{amssymb}
\usepackage{graphicx}
\graphicspath{{figures/}}
\usepackage{amsmath}
\usepackage{subfigure}
\usepackage{geometry}
\usepackage{subfig}
\usepackage{caption}
\usepackage{multirow}
\usepackage{booktabs}
\usepackage{tabularx}
\usepackage{bm}
\usepackage{color}
\newcommand{\tabincell}[2]{\begin{tabular}{@{}#1@{}}#2\end{tabular}} 

\journal{}

\begin{document}

\begin{frontmatter}



\title{Simulation of tumor ablation in hyperthermia cancer treatment: A parametric study}

\author[label1]{Qian Jiang}
\author[label1,label2]{Feng Ren}
\author[label1]{Chenglei Wang}
\author[label1]{Zhaokun Wang}
\author[label3]{Gholamreza Kefayati}
\author[label4]{Sasa Kenjeres}
\author[label5]{Kambiz Vafai}
\author[label1]{Yang Liu}
\author[label1]{Hui Tang\corref{cor1}}
\cortext[cor1]{Email: h.tang@polyu.edu.hk}

\address[label1]{Department of Mechanical Engineering, The Hong Kong Polytechnic University, Hong Kong, China}
\address[label2]{School of Marine Science and Technology, Northwestern Polytechnical University, Xi’an, Shaanxi 710072, China}
\address[label3]{School of Engineering, University of Tasmania, Hobart 7001, Tasmania, Australia}
\address[label4]{Transport Phenomena Section, Department of Chemical Engineering, Faculty of Applied Sciences and J. M. Burgers Center for Fluid Mechanics, Delft University of Technology, Van der Maasweg 9, Delft 2629 HZ, The Netherlands}
\address[label5]{Mechanical Engineering Department, University of California, Riverside, California 92521, USA}

\begin{abstract}
A holistic simulation framework is established on magnetic hyperthermia modeling to solve the treatment process of tumor, which is surrounded by a healthy tissue block. The interstitial tissue fluid, MNP distribution, temperature profile, and nanofluids are involved in the simulation. Study evaluates the cancer treatment efficacy by cumulative-equivalent-minutes-at-43$^\circ$C (CEM43), a widely accepted thermal dose coming from the cell death curve. Results are separated into the conditions of with or without gravity effect in the computational domain, where two baseline case are investigated and compared. An optimal treatment time 46.55 min happens in the baseline case without gravity, but the situation deteriorates with gravity effect where the time for totally killing tumor cells prolongs 36.11\% and meanwhile causing 21.32\% for $R_{CEM43}$ in healthy tissue. For the cases without gravity, parameter study of Lewis number $Le$ and Heat source number $Q_0$ are conducted and the variation of optimal treatment time are both fitting to the inverse functions. For the case considering the gravity, parameters Buoyancy ratio $N$ and Darcy ratio $R_{Da}$ are investigated and their influence on totally killing tumor cells and the injury on healthy tissue are matching with the parabolic functions. The results are beneficial to the prediction of various conditions, and provides useful guide to the magnetic hyperthermia treatment.

\end{abstract}




\end{frontmatter}


\section{Introduction}
Hyperthermia treatment, also named thermaltherapy, is a cancer therapeutic treatment procedure emerging in recent decades in which tumor tissues are locally heated to approximately above $43^{\circ}$C \cite{sharma2019nanoparticles,vilas2020magnetic,ma2019theoretical}. With the potential of only heating the tumor cells to death but protecting the surrounding healthy tissue, hyperthermia can effectively avoid side effects caused by conventional treatment methods and therefore alleviate suffering. Magnetic hyperthermia is one of hyperthermia modalities, by injecting the magnetic nanoparticles (MNPs) into tumor tissue region and exposing them to the high frequency alternating magnetic field (AMF) to locally heat the tumor cells to the appropriate temperature to ablate them \cite{perigo2015fundamentals,jose2020magnetic}. Actually, Magnetite ($\text{Fe}_3\text{O}_4$) is popularly chosen as an ideal MNPs candidate in many studies, since such iron-oxide nanoparticles bear favorable magnetic properties and low toxicity \cite{karponis2016arsenal, kosari2021transport, chang2018biologically}. The size  determines the heat induced by MNPs is only related to the relaxation losses when exposed to the AMF, and amount of heat is defined by Rosensweig's model \cite{suto2009heat,rosensweig2002heating}, which is highly related to the strength and frequency of AMF.

Owing to the difficulties on accurately predict the temperature distribution spatially and temporally, the reliable modeling on magnetic hyperthermia is a challenge \cite{raouf2020review}. Although massive numerical attempts have been conducted, there still lacks the holistic simulation framework involving the enough main factors in practice. The most popular used numerical model is Pennes's bio-beat transfer equation (PBHTE), which was proposed by Pennes in 1948, according to laboratory observations of human muscle. This model is based on thermal energy balance with consideration with heat convection of blood perfusion and heat generation induced by MNPs. Despite its easy implementation and widely application upon magnetic hyperthermia treatment predicting studies \cite{singh2020computational, mahmoudi2018magnetic, raouf2020review}, PBHTE is just an energy equation ignoring the tissue flow and mass transfer of the MNP in practice. Therefore, for further improving the accuracy of simulation tool, some studies try to involve more elements for magnetic hyperthermia treatment.

MNP concentration dominants the distribution of heat source \cite{maier2011efficacy}. Some works involve MNP mass transfer in the model of PBHTE, since it is really essential to the therapies \cite{dahaghin2021numerical, golneshan2011diffusion}. The MNP transfer tissue is defined in \cite{nicholson2001diffusion}, which has also been applied in many related studies \cite{roustaei2022effect,ooi2017mass}. Among them, Soltani et al. \cite{soltani2020effects} discussed the influence of MNP transfer, compared the temperature profile at different MNP diffusion time, and concluded that diffusion of MNP decreases the maximum temperature but expends the ablation region in a solid tumor. But as Salloum et al. \cite{salloum2008controlling} confirmed in experiment, MNP distribution can be controlled at the beginning if injection flow rate is slow enough.

Interstitial tissue flow also plays an important role during the treatment. Tang et al. \cite{tang2018impact, tang2020effect} considered the interstitial flow field by Brinkman equation, showing that velocity of interstitial tissue flow affects the distribution of MNP concentration and thereby the temperature profile. Similar studies were also conducted in References \cite{erbertseder2012coupled, astefanoaei2016thermofluid, zakariapour2017numerical} using Darcy equation, wherein Tang et al. \cite{tang2023backflow} is one of the rare cases conducting the study on the magnetic hyperthermia involving the factors of interstitial tissue flow, MNP distribution and heat transfer. These investigations unveiled the significant influence from interstitial flow field on the treatment efficacy. For the model on interstitial tissue flow, there gave a comparison in \cite{pedersen2007effects}, indicating that Darcy or Brinkman equation fails in accurate simulation in such porous media, as they cannot correctly capture microscopic changes in shear stress, while Navier-Stokes equation presents more velocity details. However, Navier-Stokes equation barely appears in interstitial flow simulation to tackle with magnetic hyperthermia problems. 

In practice, the density of MNP is several times larger than tissue flow \cite{zhang2008lattice, gibanov2017convective}, and meanwhile, temperature difference also affects the local density \cite{liu2014multiple}, so the gravity effect is significant in the real treatment environment \cite{yu2012review, soares2016iron}. The existence of gravity drives the motion of tissue flow and then influences the MNP distribution and temperature profile in healthy tissue and tumor, and thereby affects the treatment efficacy. Additionally, when MNPs move in tissue when exposed to the uniform AMF, Lorentz force arises to prevent its movement \cite{tzirtzilakis2005mathematical}. Gravity and Lorentz force are both external body force, while in simulation few studies consider them when modelling magnetic hyperthermia. 

Essentially, ablating on cells is a time-temperature combination work \cite{dewhirst2003basic}. This is critically important in cell death evaluation during magnetic hyperthermia treatment since the distribution of temperature is non-uniform spatially and transient temporally. CEM43 is equilibrium accumulated exposure time at $43^\circ C$ with the consideration of both temperature and duration, which was first proposed by Sapareto and Dewey \cite{sapareto1984thermal} and then has been well applied on cell killing in several investigations \cite{spirou2018magnetic,kandala2018temperature, singh2020computational}

Therefore, Upon the above studies, there lacks an accurate simulation tool of magnetic hyperthermia with involving interstitial flow, heat and mass transfer process, as well as the consideration of important external forces. In this study, a simulation framework is established to fill this gap by using Navier-Stokes equation with porous media and the mentioned external force on interstitial flow field, using energy equation with heat generation from MNP and heat transfer by blood perfusion on temperature field, and using concentration equation with convection and diffusion on MNP mass field. Based on this framework, parametric studies are conducted where Two situations are included: with and without gravity effect. Study investigates the influence from gravity, and reveal the treatment efficacy with two parameters of each situation. They are namely Lewis number and heat source number for cases without gravity and buoyancy ratio and Darcy ratio for the cases with gravity. CEM43 is adopted as the criterion of ablation in tumor and healthy tissue. Results of this study will provide the meaningful suggestions on magnetic hyperthermia treatment.

\section{Methodology}
\subsection{Physical model}

\begin{figure}
    \centering
    \includegraphics[width=0.6\textwidth]{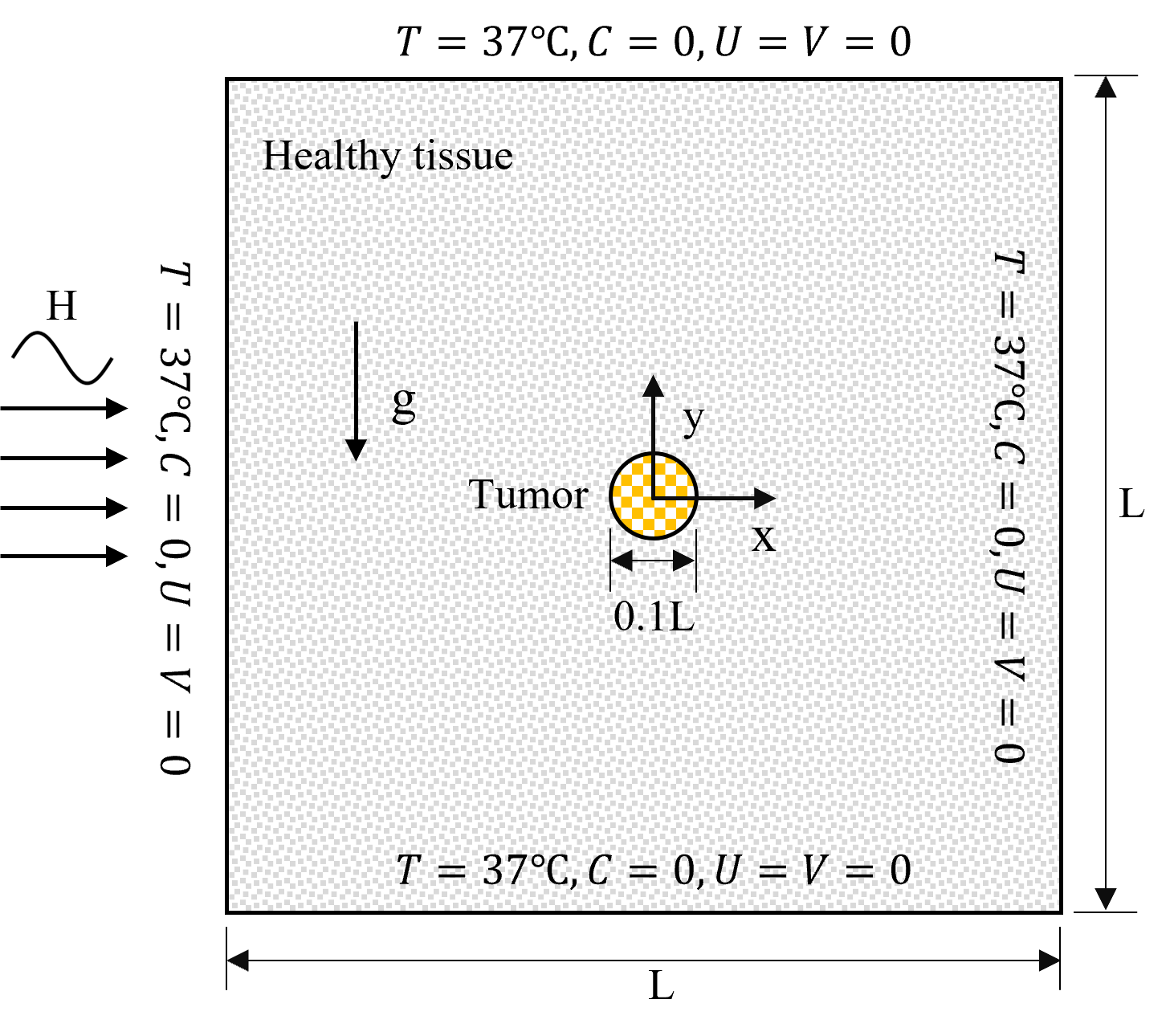}
    \caption{Schematics of magnetic hyperthermia in this study. The orange region denotes the tumor, while the gray region represents the healthy tissue. The pink part is the blood vessel.}
    \label{FigSchematics}
\end{figure}

The schematics of this problem is presented as Figure \ref{FigSchematics}, where a simplified circular tumor is located at the center of the tissue block. Diameter of tumor is assumed 10 mm, which is one-tenth of the length of whole block. Both tumor and healthy tissue are treated as porous media but with the different permeability. A straight blood vessel traverses the healthy tissue at the place near tumor with the parabolic velocity profile at inlet and open boundary at outlet. Apart from these two parts, other outer boundaries and vessel walls in fluid field are stationary. Constants zero volume fraction and core temperature ($37^\circ$C) are namely at the MNP and temperature boundaries. At the beginning of treatment, MNPs are uniformly distributed with 0.01 volume fraction in the tumor region, while the velocity of fluid is stationary and temperature is $37^\circ$C.

\subsection{Governing equation}
The governing equations of this multi-physics problem is given by

\begin{subequations}
    \begin{equation}
        \nabla\cdot\mathbf{u}=0
        \label{EqnContinuity}
    \end{equation}
    \begin{equation}
        \frac{\partial\mathbf{u}}{\partial t}+(\mathbf{u}\cdot\nabla)\Big(\frac{\mathbf{u}}{\phi}\Big)=-\frac{1}{\rho_{nf}}\nabla(\phi p)+\upsilon_{nf}\nabla^2\mathbf{u}+\frac{\mathbf{F}}{\rho_{nf}}
        \label{EqnMomentum}
    \end{equation}
    \begin{equation}
        \sigma\frac{\partial T}{\partial t}+\mathbf{u}\cdot\nabla T=\alpha_e\nabla^2T+\frac{1}{(\rho c_p)_{nf}}\dot{m}_bc_{pb}(T_b-T)+C\frac{Q}{(\rho c_p)_{nf}}
        \label{EqnEnergy}
    \end{equation}
    \begin{equation}
        \phi\frac{\partial C}{\partial t}+\mathbf{u}\cdot\nabla C=D_e\nabla^2C
        \label{EqnConcentration}
    \end{equation}
    \label{EqnGoverning}
\end{subequations}
where $\nabla\equiv\frac{\partial}{\partial x}\bm{i}+\frac{\partial}{\partial y}\bm{j}$. 

Equation \ref{EqnGoverning} namely lists continuity equation, momentum equation, energy equation, and concentration equation. In this holistic simulation framework, they are solved by multiple-relaxation-time Lattice Boltzmann Method (MRT-LBM), with the D2Q9 scheme on fluid field while D2Q5 scheme on temperature and concentration fields \cite{liu2014multiple}. $\mathbf{u}$, $T$ and $C$ denote the fluid velocity vector (u,v) along x and y directions, temperature and MNPs volume fraction, respectively. This set of equations are adaptable in tumor tissue, healthy tissue and the blood vessel region. $\phi$ means the porosity of tissue, and $p$ means pressure. $\rho$, $c_p$, $\upsilon$ are fluid density, specific heat and kinetic viscosity. Coefficient $\sigma=[(\rho c_p)_{nf}\phi+(\rho c_p)_s(1-\phi)]/(\rho c_p)_{nf}$. $k_e=k_{nf}\phi+k_s(1-\phi)$ \cite{mehmood2017numerical, hussain2018numerical} is the effective thermal conductivity of porous media, and effective thermal diffusivity is defined as $\alpha_e=k_e/(\rho c_p)_{nf}$. $D_e=\phi D$ \cite{hussain2018numerical} is the effective concentration diffusivity of MNPs in porous media, and $D$ is the concentration diffusivity in pure fluid. Specifically, subscript “$nf$” denotes properties of nanofluid or modified tissue that is a mixture combining the tissue and the injected MNPs, while “$b$” denotes properties of blood, and “$s$” denotes property of tissue structure. 

$\mathbf{F}$ in the last term of momentum equation (see Equation \ref{EqnMomentum}) is the total external body force, including the resistance force from porous media, gravity force caused by temperature and concentration gradient and Lorentz force induced by AMF as \cite{liu2014multiple}

\begin{equation}
    \mathbf{F}=-\frac{\phi\mu_{nf}}{K}\mathbf{u}-\frac{\phi\rho_{nf} F_\phi}{\sqrt{K}}|\mathbf{u}|\mathbf{u}+\phi\mathbf{G}+\phi\mathbf{F_M}
    \label{EqnExterF}
\end{equation}

where the resistance force in porous media comes from the representative elementary volume (REV) scale method \cite{guo2002lattice,liu2014multiple}. $F_\phi$ in Equation \ref{EqnGoverning} denotes the Forchheimer coefficient of porous media while $K$ denotes the permeability. $F_\phi$ is only determined by the porosity $\phi$ as $F_\phi=1.75/\sqrt{150\phi^3}$, but $K$ is determined by the combination of $\phi$ and mean pore diameter $d_p$ as $K=(\phi^3d_p^2)/[150(1-\phi)^2]$. Due to the physical difference between tumor and healthy tissue, the values of permeability are not the same in their regions, and they can be separated by the subscription as $K_{tum}$ and $K_{tis}$. $|\mathbf{u}|=\sqrt{u^2+v^2}$ is the amplitude of velocity. 

With the assumption of Boussinesq approximation, buoyancy force $\mathbf{G}$ is given by \cite{liu2018multiple}

\begin{equation}
    \mathbf{G}=g[(\rho\beta_T )_{nf}(T-T_c)+(\rho\beta_C )_{nf}(C-C_c )]\mathbf{j}
    \label{EqnExterG}
\end{equation}

where $g$ is the acceleration of gravity. $(\rho\beta_T )_{nf}$ and $(\rho\beta_C )_{nf}$ are the thermal and concentration expansion of nanofluid respectively. $T_c$ and $C_c$ are namely the reference the temperature and concentration. In this study, $T_c$ equals to the core temperature of human body $37^\circ$C, and $C_c$ is zero volume fraction. $\mathbf{j}$ is the unit vector on y direction. 

$\mathbf{F_M}$ is the Lorentz force that induced by a horizontal high frequency alternating magnetic field, which is converted into a steady model as (see \ref{AppendixLorentzforce} for detailed deviation)
\begin{equation}
    \mathbf{F_M}=-\frac{1}{2}\sigma_{nf}B_0^2v\mathbf{j}
\end{equation}
where $\sigma_{nf}$ is the electrical conductivity of nanofluid, and $B_0$ is magnetic induction amplitude, which is proportional to the magnetic field intensity amplitude $H_0$ as $B_0=\mu_0H_0$, and $\mu_0$ is magnetic permeability of vacuum. 

In energy equation (see Equation \ref{EqnEnergy}), the heat sink caused by blood perfusion $\dot m_b c_{pb}(T_b-T)$ and the heat source induced by the MNP exposed in alternating magnetic field $CQ$ are considered. $T_b=37^\circ$C is the temperature of the blood and $\dot m_b$ is the density flow rate of temperature-dependent perfusing blood. According to Lang \cite{lang1999impact}, $\dot m_b$ depends on local temperature $T$ as 

In healthy tissue:

\begin{equation}
    \dot m_b=
    \begin{cases}
            0.45+3.55 \exp [-{(T-45.0)^2}/{12.0}] & {T \le 45.0^{\circ}C} \\
            4.0  & {T > 45.0^{\circ}C} \\
    \end{cases}
\end{equation}

In tumor:

\begin{equation}
    \dot m_b=
    \begin{cases}
        0.833  & {T < 37.0^{\circ}C}\\
        0.8333- (T-37.0)^{4.8}/{5438.0} & {37.0 < T \le 42.0^{\circ}C}\\
        0.416 & {T > 42.0^{\circ}C}
    \end{cases}
\end{equation}

$Q$ is defined by the Rosensweig’s model \cite{rosensweig2002heating} as

\begin{equation}
    Q=\pi\mu_{0}\chi_{0}H^2_0f\frac{2\pi f\tau_{R}}{1+(2\pi f\tau_{R})^2}.
    \label{EqnQ}
\end{equation}

$H_0$ and $f$ are amplitude and frequency for external alternating magnetic field. $\chi_{0}$ denotes equilibrium susceptibility and $\tau_{R}$ denotes the effective relaxation time, which is determined by both Neel and Brownian relaxation time \cite{chang2018biologically}.

The following parameters are used to nondimensionalize the governing equation,
\begin{equation}
    \begin{split}
    X=\frac{x}{L},\;Y=\frac{y}{L},\; U=\frac{uL}{\alpha_f},\;V=\frac{vL}{\alpha_f},\;\tau=\frac{t\alpha_f}{L^2},\\
    P=\frac{pL^2}{\rho_f\alpha_f^2},\;\theta=\frac{T-T_c}{T_h-T_c},\; \varphi=\frac{C-C_c}{C_h-C_c}
    \end{split}
\end{equation}

Then the dimensionless governing equation is given by
\begin{subequations}
    \begin{equation}
        \nabla\cdot\mathbf{U}=0
    \end{equation}
    \begin{equation}
    \begin{split}
        &\frac{\partial\mathbf{U}}{\partial\tau}+(\mathbf{U}\cdot\nabla^*)\Big(\frac{\mathbf{U}}{\phi}\Big) \\
        &=-\frac{\rho_f}{\rho_{nf}}\nabla^*(\phi P)+Pr\frac{\upsilon_{nf}}{\upsilon_f}\nabla^{*2}\mathbf{U}-\phi\frac{\upsilon_{nf}}{\upsilon_f}\frac{Pr}{Da}\mathbf{U}-\phi\frac{F_\phi}{\sqrt{Da}}\sqrt{|\mathbf{U}|}\mathbf{U} \\
        &+[\phi\frac{(\rho\beta)_{nf}}{(\rho\beta)_f}\frac{\rho_f}{\rho_{nf}}RaPr(\theta+N\varphi)-\phi\frac{\sigma_{nf}}{\sigma_n}\frac{\rho_f}{\rho_{nf}}Ha^2Pr\mathbf{U}]\mathbf{j}
    \end{split}
    \end{equation}
    \begin{equation}
        \sigma\frac{\partial \theta}{\partial\tau}+\mathbf{U}\cdot\nabla^*\theta=\frac{\alpha_e}{\alpha_f}\nabla^{*2}\theta-\frac{(\rho c_p)_f}{(\rho c_p)_{nf}}Pe\theta+\varphi\frac{(\rho c_p)_f}{(\rho c_p)_{nf}}Q_0
    \end{equation}
    \begin{equation}
        \phi\frac{\partial\varphi}{\partial\tau}+\mathbf{U}\cdot\nabla^*\varphi=\frac{\phi}{Le}\nabla^{*2}\varphi
    \end{equation}
    \label{EqnDimenLessGoverningEq3Field}
\end{subequations}

From the above dimensionless governing equations, this problem is characterized by the following dimensionless parameters:

\begin{subequations}
    \begin{equation}
        Pr=\frac{\upsilon_f}{\alpha_f}
    \end{equation}
    \begin{equation}
        Le=\frac{\alpha_f}{D}
    \end{equation}
    \begin{equation}
        Da=\frac{K}{L^2}
    \end{equation}
    \begin{equation}
        Ra=\frac{\beta_Tg(T_h-T_c)L^3}{\upsilon_f\alpha_f}
    \end{equation}
    \begin{equation}
        N=\frac{\beta_C(C_h-C_c)}{\beta_T(T_h-Tc)}
    \end{equation}
    \begin{equation}
        Ha^2=\frac{\sigma_n\mu_0^2H_0^2L^2}{\rho_f\upsilon_f}
    \end{equation}
    \begin{equation}
        Pe=\frac{\dot{m}_bc_{pb}L^2}{(\rho c_p)_f\alpha_f}
    \end{equation}
    \begin{equation}
        Q_0=\frac{QL^2(C_h-C_l)}{(\rho c_p)_f\alpha_f(T_h-T_c)}
        \label{EqnQ0}
    \end{equation}
    \label{EqnGoverningEq3FieldDimensionless}
\end{subequations}
in which, the parameters $Pr$, $Le$, $Da$, $N$, $Ha$, $Pe$ and $Q_0$ are Prandtl number, Lewis number, Darcy number, buoyancy ratio, Hartmann number, Peclet number and heat source number.

Additionally, for the reason of different permeability in tumor and tissue, the Darcy ratio is 

\begin{equation}
    R_{Da}=\frac{Da_{tum}}{Da_{tis}}=\frac{K_{tum}}{K_{tis}}
\end{equation}

The effective properties of tissue fluid should be modified by considering the influence of interspersed MNPs, and they are computed from \cite{buongiorno2006convective,gibanov2017convective}
\begin{subequations}
    \begin{equation}
        \rho_{nf}=C \rho_n+(1-C)\rho_f
    \end{equation}
    \begin{equation}
        \upsilon_{nf}=\frac{\mu_f}{\rho_{nf}(1-C)^{2.5}}
    \end{equation}
    \begin{equation}
        (\rho c_p)_{nf}=C(\rho c_p)_n+(1-C)(\rho c_p)_f
    \end{equation}
    \begin{equation}
        (\rho\beta_T)_{nf}=C(\rho\beta_T)_n+(1-C)(\rho\beta_T)_f
    \end{equation}
    \begin{equation}
        k_{nf}=k_f\frac{k_n+2k_f-2C(k_f-k_n)}{k_n+2k_f+C(k_f-k_n)}
    \end{equation}
    \begin{equation}
        \sigma_{Enf}=\sigma_{Ef}\frac{\sigma_{En}+2\sigma_{Ef}-2C(\sigma_{Ef}-\sigma_{En})}{\sigma_{En}+2\sigma_{Ef}+C(\sigma_{Ef}-\sigma_{En})}
    \end{equation}
    \label{EqnNano}
\end{subequations}
where the subscript “$f$” means pure fluid and “$n$” means nanoparticles in Equation \ref{EqnDimenLessGoverningEq3Field} and Equation \ref{EqnNano}. In this study, pure fluid represents the interstitial tissue fluid and blood flow, in which distributes the MNP $Fe_3O_4$. The kinetic and thermal proprieties of them are listed as Table \ref{TableNanofluidForPorous} \cite{zhang2008lattice,gibanov2017convective,tzirtzilakis2005mathematical}.

\begin{table*}[]\centering
    \caption{Properties of nanofluid}
    \begin{tabular}{cccc}
    \toprule
    Properties for tissue & Value & Properties for MNPs & Value  \\ \midrule
    $\rho_f (kg/m^3)$ & 1052 & $\rho_n (kg/m^3)$ & 5200  \\
    $k_f (W/mK)$ & 0.5  & $k_n (W/mK)$ & 6  \\
    $c_{pf} (J/kgK)$ & 3800 & $c_{pn} (J/kgK)$ & 670  \\
    $\beta_{Tf} (1/K)$ & $2.1\times10^{-4}$ & $\beta_{Tn} (1/K)$ & $1.3\times10^{-5}$ \\
    $\sigma_{Ef} (\Omega^{-1}\cdot m^{-1})$ & 0.7 & $\sigma_{En} (\Omega^{-1}\cdot m^{-1})$ & $2.5\times 10^4$ \\
    $\mu_f (Pa\cdot s)$ & $6.92\times 10^{-4}$ & - & -  \\ \bottomrule
    \end{tabular}
    \label{TableNanofluidForPorous}
\end{table*}

\subsection{Thermal dose}
The cumulative-equivalent-minutes-at-$43^{\circ}C$ (CEM43) model is widely accepted in thermal dose evaluating, by converting the treatment to an equivalent time on $43^{\circ}C$, as 

\begin{equation}
    CEM43=\sum_{i=1}^l{C_{EM}}^{43-T_i}\delta t
    \label{EqnCEM43}
\end{equation}
where $T_i$ is the averaged temperature in $^{\circ}C$ at the $i$th time steps, $\delta t$ represents the time interval, and $l$ denotes the total number of time steps. ${C_{EM}}$ equals to 0.5 when $T_i>43^{\circ}C$ and 0.25 otherwise \cite{sapareto1984thermal}. As CEM43 achieves 60 minutes, cells are regarded totally destroyed \cite{singh2020computational,dewhirst2003basic}. Upon this hypothesis, an ablated area ratio in tumor or surrounding healthy tissue is defined for therapeutic efficacy, i.e.
\begin{equation}
    R_{CEM43}=\frac{S_{CEM43\geq60min}}{S_{tum}}
    \label{EqnCEM43Percentage}
\end{equation}

The optimal result of $R_{CEM43}$ in tumor and healthy tissue are 1 and 0, respectively, in accordance with the expected hyperthermia treatment efficacy - totally killing the tumor cells but without destroying the healthy tissue, 

\section{Results and discussion}
In this part, two situations are considered: with gravity and without gravity. The case without gravity, that reflects the horizontal section, ignores the influence form tissue flow field; while the case driven with gravity reflects the vertical section. 

There are Two baseline cases for the situations considering gravity or not, using the properties from practical problems. Dimensionless parameters that computing from the physical case \cite{tang2020effect,zhang2008lattice,alamiri2014fluid,swartz2007interstitial} are listed as Table \ref{TableDimensionlessParameters}. 

\begin{table}[]\centering
    \caption{Dimensionless parameters on baseline case}
    \begin{tabular}{cc}
    \toprule
    Parameters & Value\\ \midrule
    $Ra_T$ & $2.05\times 10^{8}$   \\
    $Pr$ & $5.26$ \\
    $Le$ & $125.08$ \\
    $N$ & $-18.78$ \\
    $Ha$ & $4.27$3 \\
    $Q$& $572.73$ \\
    $Da_{tis}$ & $2.00\times10^{-11}$ \\
    $R_{Da}$ & $4.84$ \\
    $\phi$ & $0.26$\\ \bottomrule
    \end{tabular}
    \label{TableDimensionlessParameters}
\end{table}

\subsection{Without gravity}
\begin{figure*}
    \centering
    \includegraphics[width=0.9\textwidth]{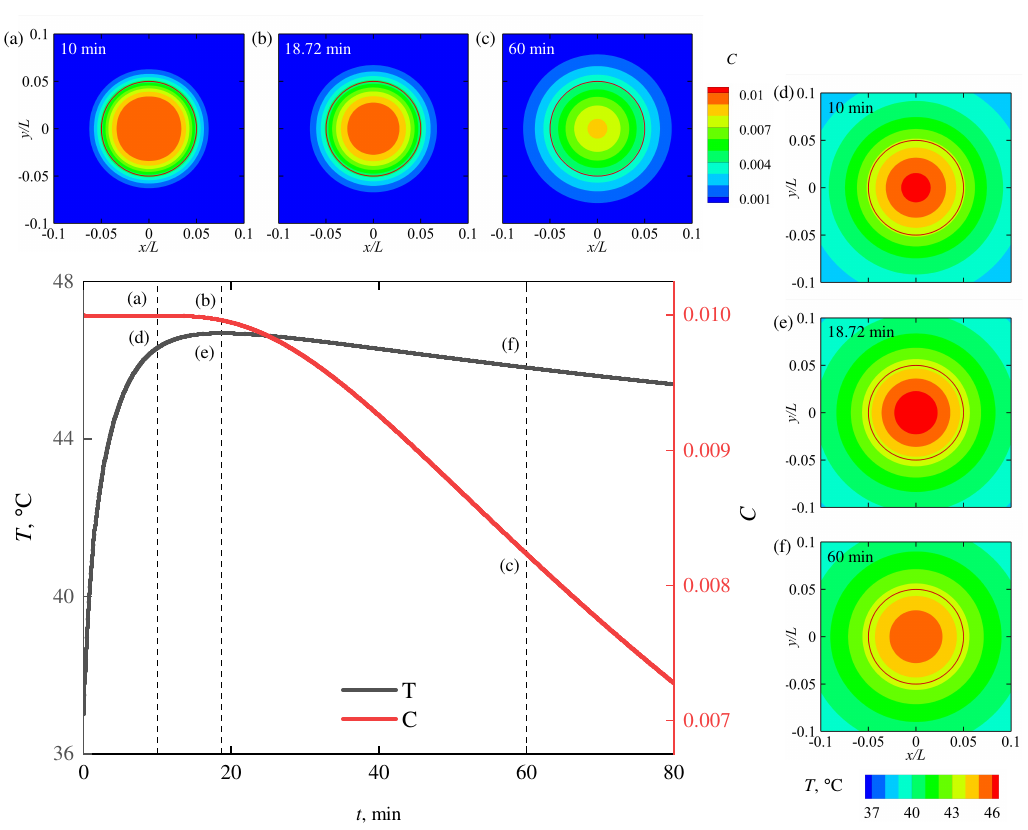}
    \caption{MNP and temperature evolution at the tumor center, with distribution at some specific treatment time for baseline case without gravity. (a)$\sim$(c) denote MNP volume fraction profiles at 10 min, 18.72 min, and 60 min, respectively; (d) $\sim$ (f) illustrate temperature profiles at 10 min, 18.72 min, and 60 min, respectively.}
    \label{FigTCCenterNoGravity}
\end{figure*}

\begin{figure*}
    \centering
    \includegraphics[width=0.9\textwidth]{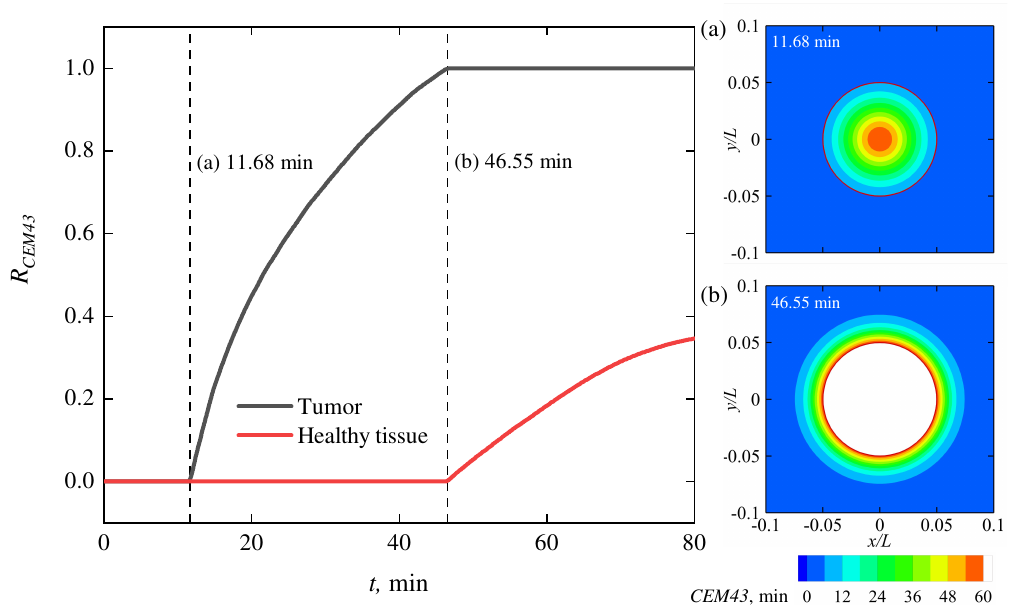}
    \caption{$R_{CEM43}$ evolution in tumor and healthy tissue for baseline case without gravity. (a) and (b) are distribution of CEM43 at 11.68 min and 46.55 min, respectively, for baseline case without gravity. White region denotes the ablation part}
    \label{FigCEM43NoGravity}
\end{figure*}

When gravity is ignored, the driven force in tissue fluid disappears, resulting in the coupling fields of temperature and MNP concentration. Figure \ref{FigTCCenterNoGravity} present MNP and temperature evolution at the tumor center, with their distribution at some typical times, for baseline case without considering gravity. Since no influence from tissue fluid, MNP diffuses with time and preforms as concentric circles on contours, then causing the concentric circles distribution on thermal field. Figure \ref{FigTCCenterNoGravity} (a)$\sim$(c) illustrate MNP profiles. As the initial condition claims, the volume fraction of MNP is 0.01 in tumor and 0 in healthy tissue. By isotropic hypothesis, MNP homogeneously spread from tumor region to the surroundings, and obviously MNP volume fraction degradation firstly occurs at the tumor boundary and gradually spread towards the tumor center. In consequence, MNP volume fraction at tumor center still remains 0.01 at 10 treatment minutes, the same as the initial level (see Point (a)). Then this concentration value slightly drop to 0.0099 at 18.72 min (see Point (c), time for highest temperature at tumor center), and finally to 0.0082 at 60 min (see Point (c)). Correspondingly, Figure \ref{FigTCCenterNoGravity} (d)$\sim$(f) reveal the thermal evolution. As $Le=125.08$, diffusion of heat is much faster than that of MNP, which leads to the results that MNP only evidently appears in the radius of 0.06 from tumor center at 10 minutes (see \ref{FigTCCenterNoGravity} (a)), but the heating region has already covered the most computation domain (see \ref{FigTCCenterNoGravity} (d)). Continuous heat from MNPs sharply increase the temperature at tumor center from $37^\circ C$ at initial time to $46.32^\circ C$ at 10 minutes (see point (d)) and then marginally increase and level off at $46.69^\circ C$ around 18.72 minutes (see point (e)). After that, the center temperature gradually decrease with MNP diffusion, finally falling to $45.82^\circ C$ at 60 minutes (see point (f)).

Figure \ref{FigCEM43NoGravity} illustrates the time history of $R_{CEM43}$ in tumor and healthy tissue. Obviously, the value of $R_{CEM43}$ in both tumor and healthy tissue are exactly both 0 at 11.68 min (see dash line (a)), where no ablation occurs in computation domain, but the CEM43 achieves 59.99 at tumor center at this time, as Figure \ref{FigCEM43NoGravity} (a). Then ablation part expends homogeneously and the best treatment efficacy merges at 46.55 min, totally killing on tumor but no injury on healthy tissue (see Figure \ref{FigCEM43NoGravity} (b)). It is noticed that the largest temperature difference in the whole tumor is $3.13^\circ C$, while leading to 34.87 s for $R_{CEM43}$ from 0 to 1. This means the critical effect of temperature fluctuation on cells killing duration, as revealed in the experiment on Chinese hamster ovary and human malignant melanoma cells \cite{roizin1991response}.

\begin{figure*}
    \centering
    \includegraphics[width=0.9\textwidth]{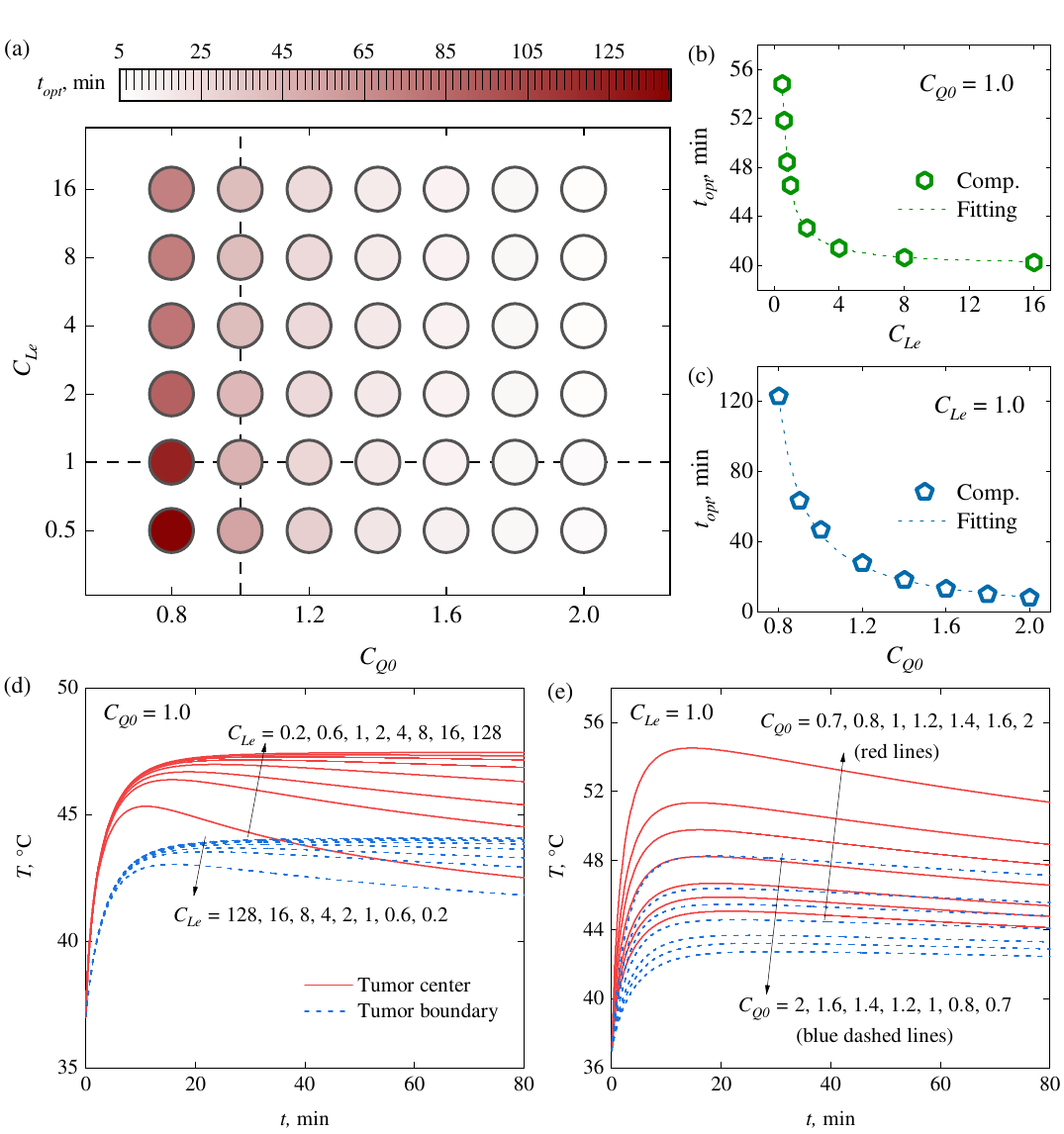}
    \caption{Effect of Lewis number $Le$ and heat source number $Q_0$ on the optimal treatment time $t_{opt}$ ((a) $\sim$ (c)) and temperature in tumor ((d) and (e)). (a) denotes $t_{opt}$ with the combination effect of $Le$ and $Q_0$, while (b) and (c) are namely $t_{opt}$ at $C_{Q0}=1.0$ and $C_{Le}=1.0$, respectively, in which the dash lines denote the corresponding fitting data in inverse functions. The fitting functions are namely (b) $t_{opt}=5.70\times(C_{Le}-0.12)^{-1}+39.97$, and (c) $t_{opt}=13.46\times(C_{Q0}-0.69)^{-1}+0.0179$. $C_{Q0}$ and $C_{Q0}$ are namely the ratio of $Le$ and $Q_0$ over baseline case. }
    \label{FigQLeTimeTem}
\end{figure*}

Lewis number $Le$ is the ratio of heat conductivity over concentration diffusivity and it varies with tissue type \cite{baxter1989transport}. Heat source number $Q_0$ reflects the capability of MNP heat generation, affected by MNP concentration, and the amplitude and frequency of alternating magnetic field \cite{raouf2020review,raouf2021parametric}. $Le$ is the internal factor while $Q_0$ is the external factor, and they are 125.08 and 572.78 respectively on baseline case.

Figure \ref{FigQLeTimeTem} (a) demonstrates the variance on the bast treatment time when $Le$ and $Q_0$ change, with highlight on $C_{Q0}$ = 1.0 (Figure \ref{FigQLeTimeTem} (b)) and $C_{Le}$ = 1.0 (Figure \ref{FigQLeTimeTem} (c)). $C_{Le}$ and $C_{Q0}$ are namely the ratio of $Le$ and $Q_0$ over baseline case. The optimal treatment time reduces monotonically with $Le$ and $Q_0$, and obviously it presents a comparatively significant change with $Q_0$ than $Le$. Take the example of $C_{Q0}$ = 1.0 and $C_{Le}$ = 1.0, similar tendency of the optimal treatment time occur - decreasing with the reduced gradient. In Figure \ref{FigQLeTimeTem} (b), variation gradient of $t_{opt}$ substantially slow down after $C_{Le}$ = 2.0. Generally, mass transfer diminishes with the Lewis number \cite{kefayati2018double, reza18-1, reza18-2} for fixed thermal diffusivity, so it is not difficult to understand that a larger Lewis number results in slower diffusion and the higher temperature in tumor. When Lewis number becomes large enough, this diffusion behavior can almost be ignored. Then $t_{opt}$ becomes the smallest value and barely changes with $Le$, which can be illustrated by $t_{opt}$ value on $C_{Le}$ = 8.0 and 16.0 (namely 40.65 min and 40.25 min, only 0.99\% difference). On the contrary, a lower $Le$ induces rapid MNP diffusion, which accelerates the temperature drop during treatment as shown Figure \ref{FigQLeTimeTem} (d). Once $Le$ is greater than a certain value, heat from MNP cannot support for killing all the tumor cells, like $C_{Le}$ = 0.2. Although a treatment lasts for 150 minutes, there still a 14\% tumor cells survive. In Figure \ref{FigQLeTimeTem} (c), $t_{opt}$ shortens in a gradually decreased gradient and still shows decline trend after $C_{Q0}$ = 2.0. Probably $t_{opt}$ approaches to 0 as $Q_0$ enlarges, but it should be noticed $Q_0$ should not be too large since the product of frequency and amplitude of magnetic field should not exceed $5\times10^9A/(ms)$ to ensure there is not obvious discomfort of patient during treatment \cite{dutz2013magnetic}. Increasing value on $Q_0$ generates more heat from MNP, considerably rising the temperature in tumor from the beginning treatment to the end when slightly elevate $Q_0$ (see Figure \ref{FigQLeTimeTem} (e)). This demonstrates the comparatively significant effect on $t_{opt}$. Furthermore, data fitting of the optimal treatment time $t_{opt}$ with variation of $C_{Le}$ and $C_{Q0}$ are respectively conducted as the dash lines in Figure \ref{FigQLeTimeTem} (b) and (c). Specially, both of them are fit to the inverse functions and the optimal treatment time $t_{opt}$ for desired values on $C_{Le}$ and $C_{Q0}$ can directly be predicted.

Table \ref{TabQLeTemNoGravity} lists the temperature at monitored points for highest value during treatment and the final computation value at 80 min. They vary with the change of $C_{Le}$ and $C_{Q0}$. Consistent with the optimal treatment results, temperature gradually reduce change with $C_{Le}$ and when $C_{Le}$ is up to 8.0, the values at both tumor center and tumor boundary are much approaching to those of $C_{Le}$ = 16.0, which are almost remain the same at one point because of the low diffusion of MNP. However, the $C_{Q0}$ appears different influence on temperature, and there shows nearly 10$^\circ C$ gap at tumor center and 10$^\circ C$ gap at tumor boundary for highest temperature. Then it is not surprising on the significant divergence of treatment time.

\begin{table*}[]
    \centering
    \caption{Temperature for highest value and final value (at 80 min) at tumor center and tumor boundary}
    \begin{tabular}{ccccccc}
    \toprule
    \multirow{8}{*}{Highest $T, ^\circ C$} & $C_{Le}$ & Tumor center & Tumor boundary & $C_{Q0}$ & Tumor center & Tumor boundary \\ \midrule
        & 0.2 & 45.33 & 43.05 & 0.7 & 45.08 & 42.74  \\ 
        & 0.6 & 46.38 & 43.54 & 0.8 & 45.90 & 43.23  \\
        & 1.0 & 46.69 & 43.69 & 1.0 & 46.69 & 43.69  \\
        & 2.0 & 46.98 & 43.84 & 1.2 & 48.25 & 44.59  \\
        & 4.0 & 47.17 & 43.93 & 1.4 & 49.80 & 45.49  \\
        & 8.0 & 47.29 & 44.00 & 1.6 & 51.36 & 46.41  \\
        & 16.0 & 47.36 & 44.04 & 2.0 & 54.53 & 48.29  \\ \midrule
    \multirow{8}{*}{Final $T, ^\circ C$}
        & 0.2 & 42.51 & 41.83 & 0.7 & 44.13 & 43.47  \\ 
        & 0.6 & 44.53 & 42.94 & 0.8 & 44.78 & 42.90  \\
        & 1.0 & 45.39 & 43.31 & 1.0 & 45.39 & 43.31  \\
        & 2.0 & 46.31 & 43.64 & 1.2 & 46.59 & 44.08  \\
        & 4.0 & 46.87 & 43.85 & 1.4 & 47.76 & 44.82  \\
        & 8.0 & 47.17 & 43.96 & 1.6 & 48.95 & 45.58  \\
        & 16.0 & 47.32 & 44.03 & 2.0 & 51.39 & 47.16 \\ \bottomrule
    \end{tabular}
    \label{TabQLeTemNoGravity}
\end{table*}

\subsection{With gravity}
\begin{figure*}
    \centering
    \includegraphics[width=\textwidth]{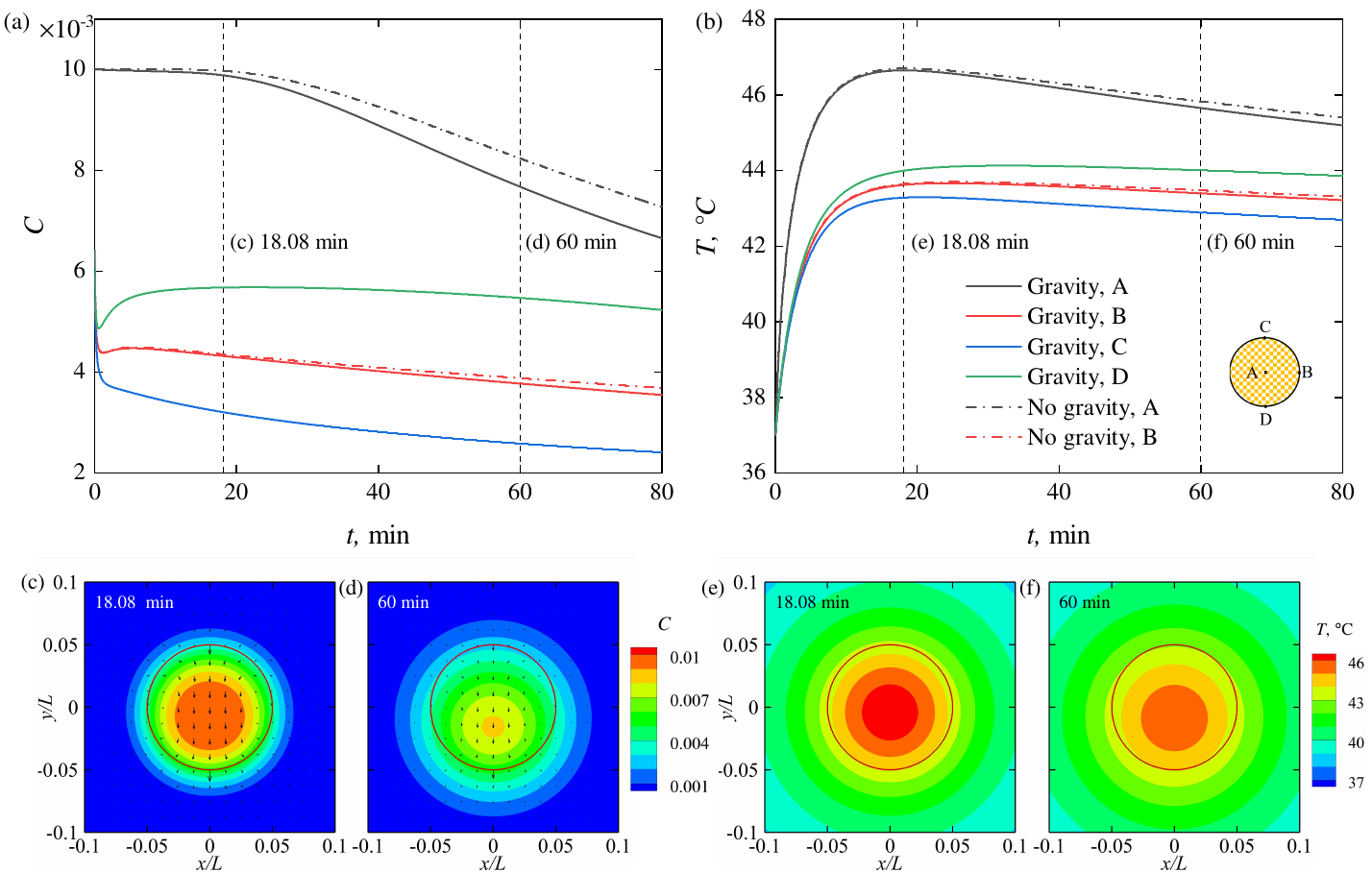}
    \caption{MNP and temperature evolution at the monitored points of tumor center (point A), left boundary (point B), upper boundary (point C) and lower boundary (point D), as well as the distribution at some specific treatment time for baseline case with gravity (compared with the case without gravity). (c) and (d) denote MNP volume fraction profiles at 18.74 min and 60 min, respectively; (e) and (f) illustrate temperature profiles at 18.74 min and 60 min, respectively. For symmetric distribution, point B represents the situations of each boundary points}
    \label{FigTCCGravity}
\end{figure*}

\begin{figure*}
    \centering
    \includegraphics[width=0.9\textwidth]{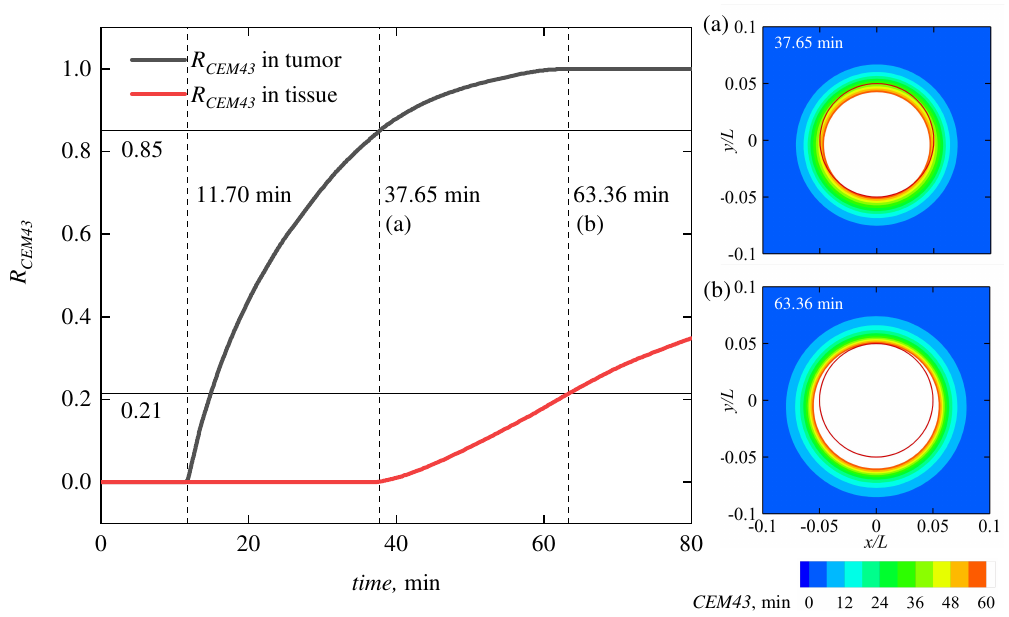}
    \caption{$R_{CEM43}$ evolution in tumor and healthy tissue for baseline case with gravity. (a) and (b) are distribution of CEM43 at 37.65 min and 63.36 min, respectively, for baseline case with gravity. White region denotes the ablation part}
    \label{FigCEM43Gravity}
\end{figure*}

Since magnetite is usually chosen as ideal MNP material \cite{chang2018biologically,kosari2021transport}, whose density is substantially higher than tumor and healthy tissue, a vertical movement tendency inevitably exists. Then the temperature distribution is coupled with the MNP volume fraction and tissue fluid flow velocity. 

Figure \ref{FigTCCGravity} (a) and (b) present the distribution of MNP and temperature evolution at the monitored points: tumor center - point A, left boundary - point B, upper boundary - point C and lower boundary - point D. Heat and mass transfer process is similar as analyzed in the case without gravity, but the influence of gravity pulls the MNP down, as well as the high-temperature region ((see Figure \ref{FigTCCGravity} (c)$\sim$(f))). Although sightly, the time for maximum temperature at tumor center is advanced to 18.08 min, and this maximum value is diminished to 46.64 $^\circ C$ compared with no gravity (see Figure \ref{FigTCCGravity} (e)), then which reduces to 45.65 $^\circ C$ at 60 min (see Figure \ref{FigTCCGravity} (f)), a gradually-increased gap with no gravity case. Such differences at four monitored points are quantified in the Table \ref{TabTCompareWithNoGravity}. Obviously, the influence on temperature enlarges with time, and gravity effect evokes more remarkable difference at upper and lower tumor boundary. Among four monitored points, only Point D benefits from the gravity, obtaining the maximum 0.54 $^\circ C$ difference during the treatment. Buoyancy ratio for this baseline case equals to -18.77 (downwards direction, opposite to y axis), where MNPs have higher density than tissue, so they drop towards the lower boundary of tumor. This movement exerts two symmetrical vortex on tissue interstitial fluid as Figure \ref{FigTCCGravity} (c) and (d), and cause high temperature region drop by convection, similar as revealed by Vijaybabu \cite{vijaybabu2021influence}. 

\begin{table*}[]
    \centering
    \caption{Temperature for highest value and final value (at 80 min) at monitored points}
    \begin{tabular}{cccccc}
    \toprule
    \multicolumn{2}{c}{Condition}                  & Point A & Point B & Point C & Point D \\ \midrule
    \multirow{2}{*}{Highest $T, ^\circ C$}   & without gravity & 46.69   & 43.69   & 43.69   & 43.69   \\ 
                                 & with gravity    & 46.64   & 43.66   & 43.29   & 44.12   \\
    \multirow{2}{*}{Final $T, ^\circ C$}     & without gravity & 45.39   & 43.31   & 43.31   & 43.31   \\
                                 & with gravity    & 45.19   & 43.21   & 42.69   & 43.85   \\ \bottomrule
    \end{tabular}
    \label{TabTCompareWithNoGravity}
\end{table*}

Figure \ref{FigCEM43Gravity} presents the time history of $R_{CEM43}$ in tumor and healthy tissue for baseline case with gravity effect. Results indicate injury on healthy tissue is inevitable during treatment on current situation. Time for totally killing tumor cells enlarges to 63.36 min, 36.11\% longer than that without gravity (see Figure \ref{FigCEM43Gravity} (a)). In the meanwhile, 21.32\% injury occurs in healthy tissue, and this injury begins from 37.65 min, when tumor killing only achieves 85.04\% (see Figure \ref{FigCEM43Gravity} (b)). Therefore, a time span exists from 37.65 to 63.36 min, during which both tumor and healthy tissue cells are killed. Therefore gravity effect deteriorates treatment efficacy in practical, and some measures need to be adopted to improve the situation.

\begin{figure*}
    \centering
    \includegraphics[width=\textwidth]{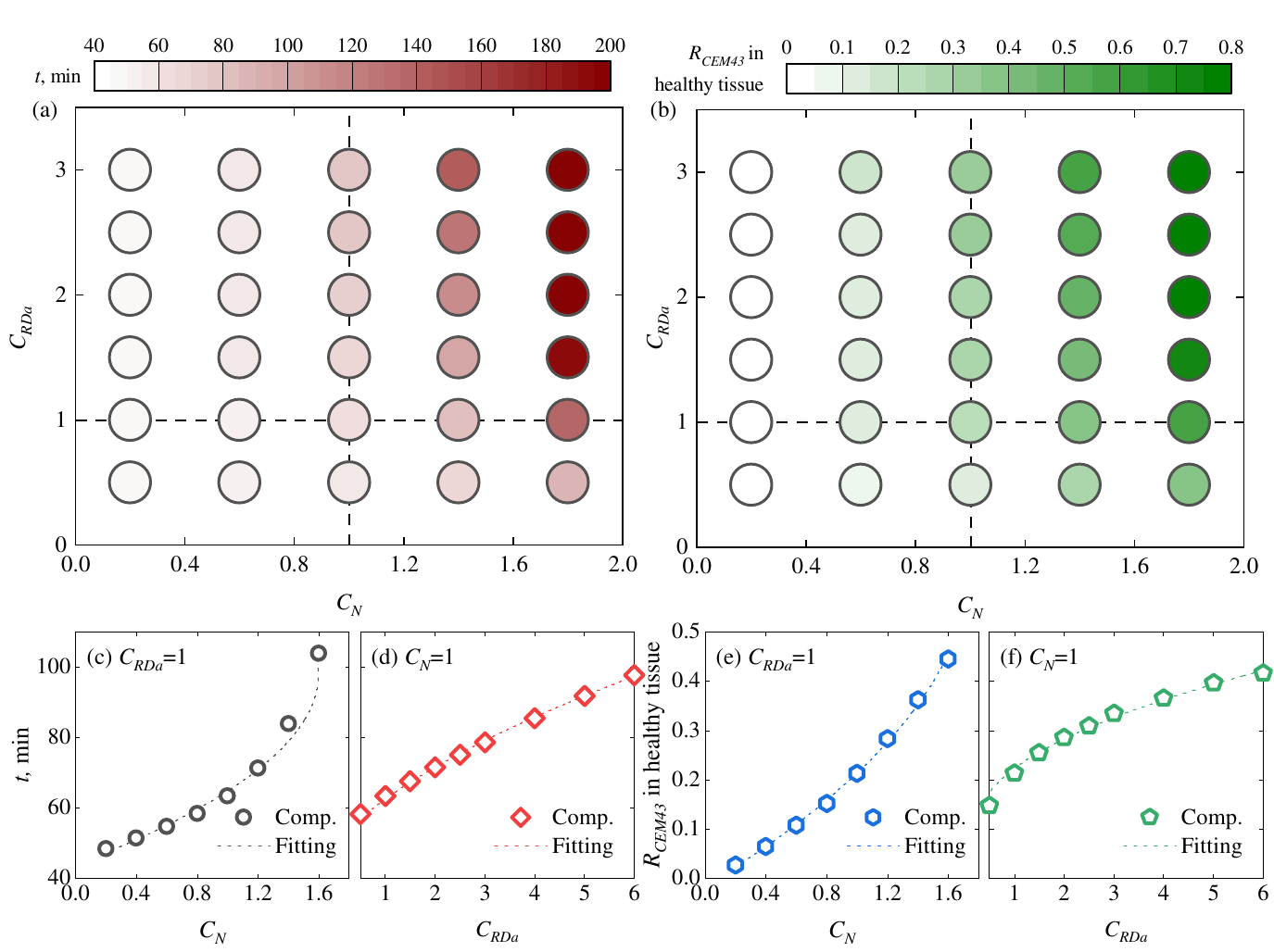}
    \caption{Effect of buoyancy ratio $N$ and Darcy ratio $R_{Da}$ on (a) treatment time $t$ of totally killing tumor cells and (b) injury $R_{CEM43}$ in healthy tissue. (c) and (d) denote $t$ at $C_N$ = 1.0 and $C_{RDa}$ = 1.0, respectively, while (e) and (f) are namely $R_{CEM43}$ in healthy tissue at $C_N$ = 1.0 and $C_{RDa}$ = 1.0, respectively. The fitting functions are parabolic, and they are namely (c) $C_N=-0.00054\times t^2+0.11\times t-3.62$, (d) $C_{RDa}=0.0013\times t^2-0.062\times t-0.15$, (e) $C_N=-3.97\times R_{CEM43}^2+5.13\times R_{CEM43}+0.083$, and (f) $C_{RDa}=72.14\times R_{CEM43}^2-21.10\times R_{CEM43}+2.12$.$C_N$ and $C_{RDa}$ are namely the ratio of $N$ and $R_{Da}$ over baseline case}
    \label{FigNRDaTimeRCEM}
\end{figure*}

Buoyancy ratio $N$ is the comparison of buoyancy force induced by concentration difference and temperature difference,while Darcy ratio $R_{Da}$ reflects the relation of permeability in tumor and healthy tissue. They are namely $N = -18.77$ (`` $-$ " means the opposite direction of buoyancy forces) and $R_{Da}$ = 4.84 in baseline case.

Figure \ref{FigNRDaTimeRCEM} illustrates the time $t$ for totally killing the tumor region and the damage $R_{CEM43}$ meanwhile in the healthy tissue, respectively, when $N$ and $R_{Da}$ change, with highlight on $C_N$ = 1.0 and $C_{RDa}$ = 1.0. It is not difficult to discover that time for totally destroying tumor cell $t$ and the injury meanwhile on healthy tissue $R_{CEM43}$ perform similar tendency - they both rise from the left lower corner to upper right corner. Take the example of $C_{RDa}=1.0$ and $C_N=1.0$, $t$ and $R_{CEM43}$ in healthy both enlarge, in an increasing gradient with $C_N$ while in a decreasing gradient with $C_{RDa}$ as Figure \ref{FigNRDaTimeRCEM} (c) $\sim$ (f). Surprisingly, they seem just scattered along the parabola equations. Therefore, with the equations $t$ and $R_{CEM43}$ in healthy tissue can be predicated as $C_N$ and $C_{RDa}$ vary.

\begin{figure*}
    \centering
    \includegraphics[width=0.95\textwidth]{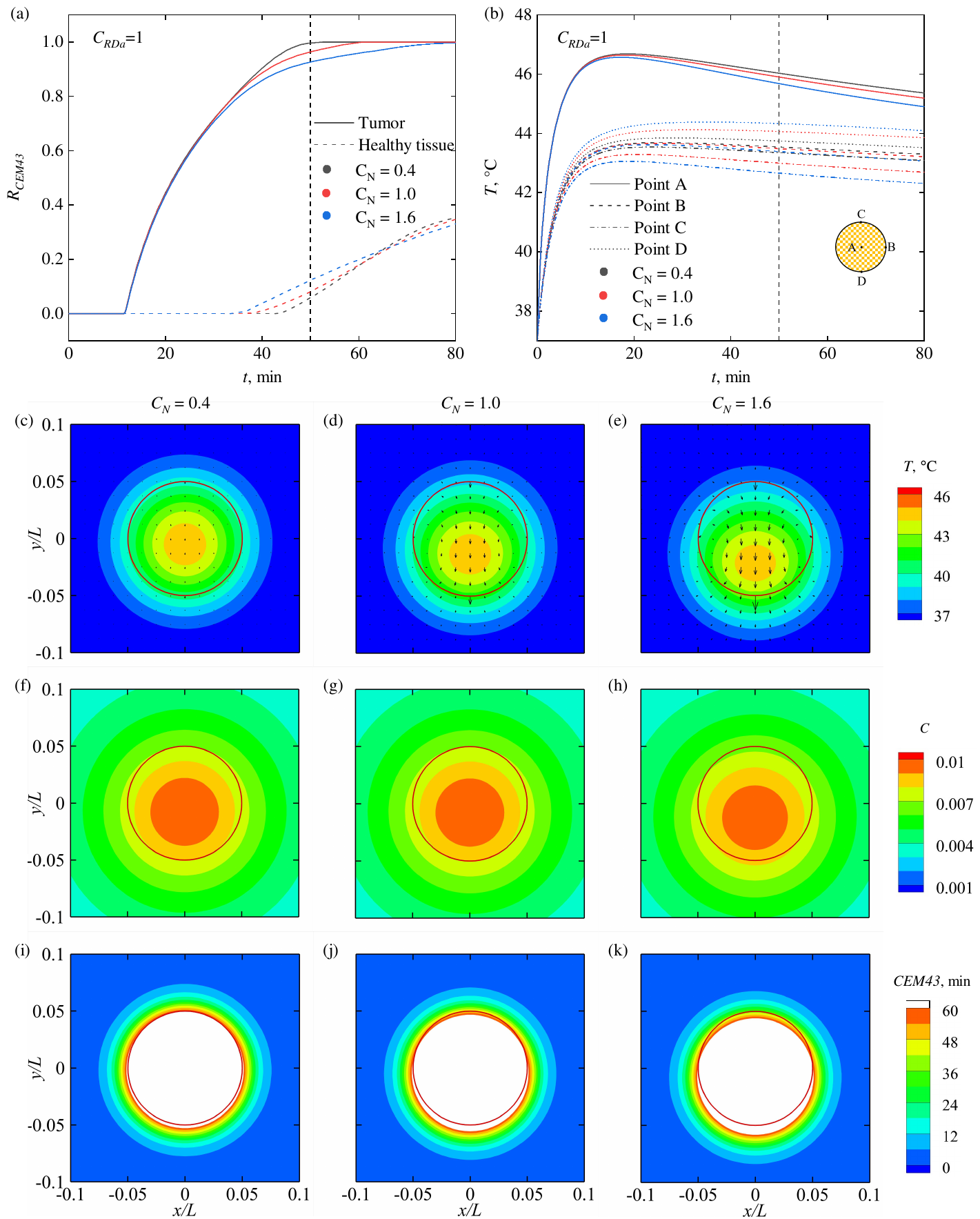}
    \caption{Evolution of $R_{CEM43}$ in both tumor and healthy tissue (a), and evolution of $T$ at monitored points (b) with effect of buoyancy ratio $N$. $C_N$ means the ratio of $N$ over the value of baseline case. (c) $\sim$ (k) are MNP concentration, temperature distribution, and CEM43 distribution for $C_N$ = 0.4, 1.0 and 1.6 at treatment time $t$ = 50 min. Arrows in (c) $\sim$ (e) denote the velocity. Red circle denotes the tumor boundary, while white region means the destroyed part.}
    \label{FigCNRCEMTC}
\end{figure*}

Figure \ref{FigCNRCEMTC} (a) demonstrates $R_{CEM43}$ time history with the variation of buoyancy ratio $N$ in both healthy and tumor region. Obviously, at the beginning of ablation in tumor region there is not considerable difference, but the increasing $C_N$ slows down the killing process on tumor cells at the end of treatment. The time for destroying all the tumor region $t$ prolongs from 51.39 min at $C_N=0.4$, to 103.99 min at $C_N=1.6$, a value of more than 2 times on the former.
Meanwhile, in the healthy tissue, growth on $C_N$ advance the ablation time from 43.85 min at $C_N=0.4$ to 34.24 min at $C_N=1.6$. However, this is not surprising. As the divergence on $R_{CEM43}$ and $T$ appear apparent at 50 min treatment time, the instant fields information is used as the example to explain the difference. Enlarge on $C_N$ theoretically augments the downwards buoyancy force in Equation \ref{EqnDimenLessGoverningEq3Field}b, which speeds downwards velocity near high MNP concentration region as Figure \ref{FigCNRCEMTC} (c) $\sim$ (e). Then temperature distribution is affected by the heat MNP heat source and the convection with velocity (as Equation \ref{EqnDimenLessGoverningEq3Field}c), thereby enhancing the drop of high temperature region at larger $C_N$ (see Figure \ref{FigCNRCEMTC} (f) $\sim$ (h). Temperature at Point C and Point D considerably change with $C_N$. When developed to 50 min, temperature reduces 0.69 $^\circ C$ at Point C while gathers 0.59 $^\circ C$ from $C_N$ = 0.4 to 1.6, and the difference still expends with time. Although it seems tiny, the accumulation on temperature difference significantly separates the ablation process at the upper and lower tumor boundary as Figure \ref{FigCNRCEMTC} (i) $\sim$ (k), where almost all the tumor cells are destroyed with little accidental injury on healthy tissue, but the situation deteriorates with $C_N$, therefore leading to a huge gap on treatment time $t$.

\begin{figure*}
    \centering
    \includegraphics[width=0.95\textwidth]{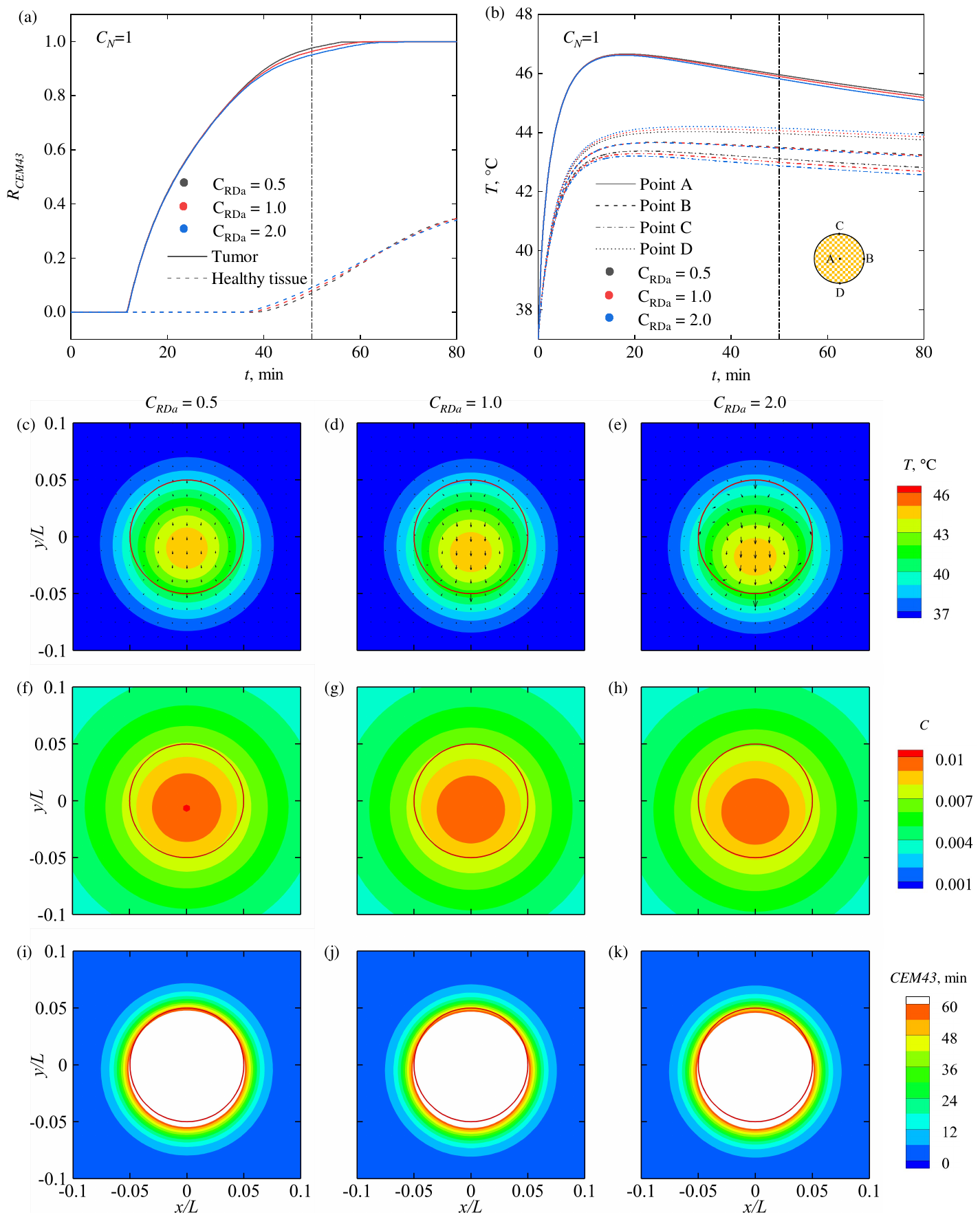}
    \caption{Evolution of $R_{CEM43}$ in both tumor and healthy tissue (a), and evolution of $T$ at monitored points (b) with effect of buoyancy ratio $R_{Da}$. $C_{RDa}$ means the ratio of $R_{Da}$ over the value of baseline case. (c) $\sim$ (k) are MNP concentration, temperature distribution, and CEM43 distribution for $C_{RDa}$ = 0.5, 1.0 and 2.0 at treatment time $t$ = 50 min. Arrows in (c) $\sim$ (e) denote the velocity. Red circle denotes the tumor boundary, while white region means the destroyed part.}
    \label{FigCRDaRCEMTC}
\end{figure*}

Similarly, Figure \ref{FigCRDaRCEMTC} illustrates the situation with the change on Darcy ratio $R_{Da}$. which exits in the resistant force induced by porous media. $C_{RDa}$ means the ratio of $R_{Da}$ over the value of baseline case. The augment on Darcy ratio $C_{RDa}$ means the resistant force on velocity becomes weaken in tumor region (as Equation \ref{EqnDimenLessGoverningEq3Field}b) so that the downwards velocity also enlarges as Figure \ref{FigCRDaRCEMTC} (c) $\sim$ (e). Therefore, the MNP and temperature show more downwards tendency, then prolonging the tumor treatment time $t$ and advancing the healthy tissue injury. However, the change on treatment time in tumor region and ablation time in healthy tissue with $C_{RDa}$ are not much significant as occurred with $C_N$. They are 13.27 min delay and 3.11 min advance, respectively.

\section{Conclusion}
A holistic MRT-LBM simulation framework on the magnetic hyperthermia treatment is established, with the multiphysics of porous flow, heat and mass transfer, nanofluids involved. Simulation reveal the distribution of MNP concentration, interstitial tissue fluid flow and temperature in tumor and healthy tissue, and the treatment efficacy. Among them, treatment efficacy is evaluated by CEM43 model, in which the cells death is the combination outcome of the temperature and heating time. This framework is well validated in multiphysics problem and properly behaves in dealing with the magnetic hyperthermia treatment problems. Results are separated into the situations of with or without gravity include, where two baseline cases respectively and the parameters are investigated. 

For the baseline case with no gravity, tissue flow is stationary since no driving force exists. With such, contours for both MNP concentration and temperature profile are concentric circle shapes. Consequently, an optimal treatment time 46.55 min happens by which tumor cells are totally destroyed but no injury occurs on surroundings healthy tissue. Lewis number $Le$ directly affects the MNP diffusion rate, which then changes the temperature, especially at later treatment stage. The larger the $Le$, the more the temperature drop, the longer the optimal treatment time. Heat source number $Q_0$ influence the amount of heat from unit dose of MNP. Larger $Q_0$ lead to the reasonable temperature climb in the entire treatment process, and significantly reduce the optimal treatment time. 

For the baseline case considering gravity effect, MNP concentration and temperature differences influence the gravity force, and induce downwards drop of fluid and then MNPs and temperature. This prolongs the time for totally killing the tumor region with 36.11\% increment and cause 21.32\% injury on healthy tissue. Enlarge on the value of Buoyancy ratio $N$ increases the downwards gravity force and enhances the MNP drop, which substantially slow down the tumor ablation at last stage and noticeably advance the injury on healthy tissue. Increase on Darcy ratio $R_{Da}$ reduce the porous resistance force, and presents the similar tendency on cell killing process as $N$ but a smaller variation. 

Variation of Lewis number $Le$ and Heat source number $Q_0$ on the optimal treatment time $t_{opt}$ are fitting as inverse functions, while the influence from Buoyancy ratio $N$ and Darcy ratio $R_{Da}$ are matching with the parabolic functions. These functions are beneficial to the prediction of various conditions, and the results can provide useful guide to the magnetic hyperthermia treatment.

\section*{Acknowledgement}
This study was financially supported by the Research Grants Council of Hong Kong under General Research Fund (Project No. 15214418).

\appendix
\section{Lattice Boltzmann method}
The LBM is a relatively new CFD method for fluid flow and heat/mass transfer simulations. Unlike traditional CFD methods, which solve the conservation equations of macroscopic properties numerically, LBM models the fluid particles by distribution functions through consecutive streaming and collision processes over a number of square lattices \cite{wang2022simulation,ren19,wang2016control}. Zhang \cite{zhang2008lattice} was probably the first to apply the LBM to solving PBHTE, successfully demonstrating the capability of LBM in simulating bioheat problems. This mesoscopic approach was then widely applied for bioheat studies \cite{golneshan2011diffusion,das2013estimation,das2013numerical}. In the present study, a D2Q9 (i.e., two-dimensional nine-discrete-velocity) MRT (i.e., multiple-relaxation-time, a collision model that is used to improve the numerical stability \cite{lallemand2000theory}) and a D2Q5 (i.e., two-dimensional five-discrete-velocity) MRT LBM scheme are namely applied to obtain the vector field (flow field)  and scalar field (temperature field and concentration field) by solving Equation \ref{EqnGoverning} \cite{liu2014multiple,liu2018multiple}.

The discrete D2Q9 MRT-LBM equation for velocity field is written as
\begin{equation}
    \mathbf{f}(x_k+\mathbf{e}\delta_t,t_n+\delta_t)-\mathbf{f}(x_k,t_n)=-\mathbf{M}^{-1} \mathbf{\Lambda}[\mathbf{m}-\mathbf{m}^{(eq)}]|_{(x_k,t_n)}+\mathbf{M}^{-1}\delta_t(1-\frac{\mathbf{\Lambda}}{2})\mathbf{S}
\end{equation}
where $\mathbf{f}(x_k,t_n)$ is nine-dimensional distribution function vectors at time ${t}_n$ and node $x_k$ for fluid field. $\mathbf{m}$ and $\mathbf{m}^{(eq)}$ are moment and the corresponding equilibrium moment vector for flow field. $\mathbf{e}$ describes unit velocities along 9 discrete directions
\begin{equation}
    e_i=
    \begin{cases}
            (0,0) & {i=0} \\
            (\cos{[(i-1)\pi/2]},\sin{[(i-1)\pi/2]})c  & {i=1\sim 4} \\
            (\cos{[(2i-9)\pi/4]},\sin{[(2i-9)\pi/4]})\sqrt{2}c  & {i=5\sim 8}
    \end{cases}
\end{equation}
where $c=\delta_x/\delta_t$ is the lattice speed, which is 1 since $\delta_x=\delta_t$ in the MRT model. 
$\mathbf{M}$ is a $9\times5$ orthogonal transformation matrix
\begin{equation}
\mathbf{M}=
    \begin{pmatrix} 
    1&1&1&1&1&1&1&1&1 \\ 
    -4&-1&-1&-1&-1&2&2&2&2 \\
    4&-2&-2&-2&-2&1&1&1&1 \\
    0&1&0&-1&0&1&-1&-1&1 \\
    0&-2&0&2&0&1&-1&-1&1 \\
    0&0&1&0&-1&1&1&-1&-1 \\
    0&0&-2&0&2&1&1&-1&-1 \\
    0&1&-1&1&-1&0&0&0&0 \\
    0&0&0&0&0&1&-1&1&-1
    \end{pmatrix}
\end{equation}
$\mathbf{\Lambda}$ is the nine-dimensional diagonal relaxation matrix
\begin{equation}
    \mathbf{\Lambda}=\text{diag}(1,1.1,1.1,1,1.2,1,1.2,1/\tau_\upsilon,1/\tau_\upsilon)
\end{equation}
and $\tau_\upsilon$ can be recovered to viscosity of nanofluid in Chapman–Enskog analysis on Equation \ref{EqnGoverning} as
\begin{equation}
    \upsilon_{nf}=\sigma c_{s}^2(\tau_\upsilon-0.5)\delta_t
\end{equation}
$\mathbf{S}$ is the external force vector in the moment space, which is linked to the body force $F$ in governing equation \ref{EqnGoverning}(b).

The discrete D2Q5 MRT-LBM equation for temperature and concentration fields are written as
\begin{subequations}
    \begin{equation}
        \mathbf{g}(x_k+\mathbf{e}\delta_t,t_n+\delta_t)-\mathbf{g}(x_k,t_n)=-\mathbf{N}^{-1} \mathbf{\Theta}[\mathbf{n_g}-\mathbf{n_g}^{(eq)}]|_{(x_k,t_n)}+\mathbf{N}^{-1}\delta_t\mathbf{\Psi}
    \end{equation}
    \begin{equation}
    \mathbf{h}(x_k+\mathbf{e}\delta_t,t_n+\delta_t)-\mathbf{h}(x_k,t_n)=-\mathbf{N}^{-1} \mathbf{\Upsilon}[\mathbf{n_h}-\mathbf{n_h}^{(eq)}]|_{(x_k,t_n)}
\end{equation}
\end{subequations}
where $\mathbf{g}(x_k,t_n)$ and $\mathbf{h}(x_k,t_n)$ are five-dimensional distribution function vectors at time ${t}_n$ and node $x_k$ for temperature and concentration respectively. $\mathbf{n}$ and $\mathbf{n}^{(eq)}$ are moment and the corresponding equilibrium moment vector, respectively, where subscribe "g" represents temperature field and "h" denotes concentration field. $\mathbf{e}$ describes unit velocities along 5 discrete directions
\begin{equation}
    e_i=
    \begin{cases}
            (0,0) & {i=0} \\
            (\cos{[(i-1)\pi/2]},\sin{[(i-1)\pi/2]})c  & {i=1\sim 4} \\
    \end{cases}
\end{equation}
$\mathbf{N}$ is a $5\times5$ orthogonal transformation matrix
\begin{equation}
\mathbf{N}=
    \begin{pmatrix} 
    1&1&1&1&1 \\ 
    0&1&0&-1&0 \\
    0&0&1&0&-1 \\
    -4&1&1&1&1 \\
    0&1&-1&1&-1
    \end{pmatrix}
\end{equation}
$\mathbf{\Theta}$ and $\mathbf{\Upsilon}$ are the diagonal relaxation matrix
\begin{subequations}
    \begin{equation}
        \mathbf{\Theta}=\text{diag}(1,1/\tau_T,1/\tau_T,1.5,1.5)
    \end{equation}
    \begin{equation}
        \mathbf{\Upsilon}=\text{diag}(1,1/\tau_C,1/\tau_C,1.5,1.5)
    \end{equation}
\end{subequations}
where $\tau_T$ can be linked to effective thermal diffusivity (in temperature field) or effective concentration diffusivity (in concentration field) in Chapman–Enskog analysis on Equation \ref{EqnGoverning} as
\begin{equation}
    \alpha_e=\sigma c_{sT}^2(\tau_T-0.5)\delta_t,\;D_e=\phi c_{sT}^2(\tau_C-0.5)\delta_t
\end{equation}

$\mathbf{\Psi}$ is a heat source vector, which can is connected with the heat source $Q$ in governing equation \ref{EqnGoverning}(c). More details about the D2Q5 MRT LBM can be found in \cite{liu2014multiple,liu2018multiple}.

For the boundary conditions at four sides of healthy tissue block in this study, stationary wall is applied for fluid field, constant values are used on thermal and solutal fields, as depicted in Figure \ref{FigSchematics}. Here, halfway bounce-back is adopted for stationary wall, while anti-bounce-back scheme is adopted for constant temperature and concentration boundary. In addition, at the interface of healthy tissue and tumor, it is deemed the same velocity, same temperature and same concentration.

\begin{table*}[]
    \centering
    \caption{Sensitivity studies on computational domain size and grid number}
    \begin{tabular}{ccccccccc}
    \toprule
    \multirow{7}{*}{Domain size} & $L/d_{tum}$ & \tabincell{c}{$Time_{tot},$\\ $min$} & \tabincell{c}{$Err_{tot},$\\$ \%$} & \tabincell{c}{$time,$\\$min$}   & \tabincell{c}{$T_A,$\\$^\circ C$} & \tabincell{c}{$T_C,$\\$^\circ C$} & \tabincell{c}{$Err_A,$\\$ \%$} & \tabincell{c}{$Err_C,$\\$ \%$}\\ \cmidrule{2-9}
    & \multirow{2}{*}{5} & \multirow{2}{*}{71.00} & \multirow{2}{*}{18.57} & 30 & 46.40 & 43.56 & 0.09 & 1.28 \\
    & & & & 60 & 45.64 & 43.29 & 0.03 & 1.61\\
    & \multirow{2}{*}{10} & \multirow{2}{*}{63.36} & \multirow{2}{*}{0.12} & 30 & 46.44 & 44.12 & 0.13 & 0.13 \\  
    & & & & 60 & 45.65 & 44.00 & 0.12 & 0.12 \\
    & \multirow{2}{*}{20} & \multirow{2}{*}{63.71} & \multirow{2}{*}{-} & 30 & 46.38 & 44.06 & - &  - \\ 
    & & & & 60 & 45.59 & 43.95 & - & -\\   \midrule
    \multirow{7}{*}{Grid number} & $NX\times NY$ & \tabincell{c}{$Time_{tot},$\\ $min$} & \tabincell{c}{$Err_{tot},$\\$ \%$} & \tabincell{c}{$time,$\\$min$}   & \tabincell{c}{$T_A,$\\$^\circ C$} & \tabincell{c}{$T_C,$\\$^\circ C$} & \tabincell{c}{$Err_A,$\\$ \%$} & \tabincell{c}{$Err_C,$\\$ \%$}\\ \cmidrule{2-9}
    & \multirow{2}{*}{$200\times200$} & \multirow{2}{*}{75.13} & \multirow{2}{*}{12.06} & 30 & 46.38 & 44.32 & 1.46 & 0.44 \\
    & & & & 60 & 45.56 & 44.14 & 1.94 & 0.33\\
    & \multirow{2}{*}{$400\times400$} & \multirow{2}{*}{63.36} & \multirow{2}{*}{0.56} & 30 & 46.44 & 44.12 & 0.02 & 0.05 \\  
    & & & & 60 & 45.65 & 44.00 & 0.01 & 0.04 \\
    & \multirow{2}{*}{$600\times600$} & \multirow{2}{*}{63.28} & \multirow{2}{*}{-} & 30 & 46.44 & 44.10 & - &  - \\ 
    & & & & 60 & 45.64 & 43.98 & - & -\\   \midrule
    \bottomrule   
    \end{tabular}
    \label{TableMeshandBoundary}
\end{table*}

A sensitivity study of both computational domain size and grid number is conducted. The baseline case with gravity effect is considered as an example, where three sets of domain size and two sets of grid number are compared. Temperature values of specific points are monitored and time for totally killing the tumor cells $Time_{tot}$ are recorded, and they are Listed as Table \ref{TableMeshandBoundary}. Results suggests that the combination of $10d_{tum}\times10d_{tum}$ computational domain and $400\times400$ grid number is suitable for the study.

\begin{figure*}[htpb]
    \centering
    \includegraphics[width=\textwidth]{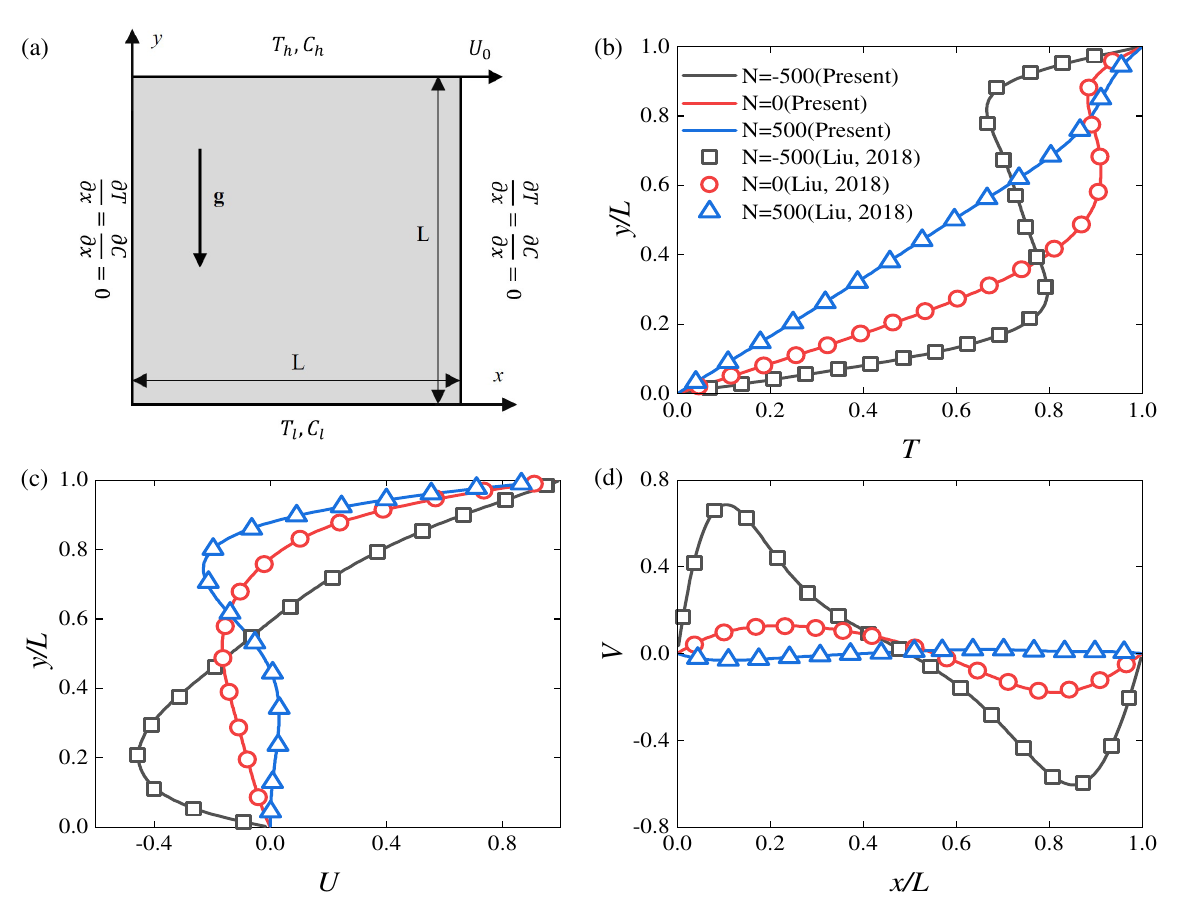}
    \caption{Validation of LBM framework on double-diffusive convection problem in porous media. (a) vertical temperature profile at x/L=0.5; (b) vertical velocity profile at x/L=0.5; (c) horizontal velocity profile at x/L=0.5.  
    (b) vertical velocity on $y=0.5$, (c) temperature on $y=0.5$}
    \label{FigDoubleDiff}
\end{figure*}

Since the magnetic hyperthermia problem governed by Equation \ref{EqnGoverning} actually can be viewed as a double-diffusive convection problem in porous media. Therefore, a typical porous double-diffusive convection validation is conducted to justify this framework. Figure \ref{FigDoubleDiff} show the comparison on velocity and temperature with Reference \cite{liu2018multiple} at various buoyancy ratio $N$. It is indicated that the present results match well with the reference.

\section{Alternating Lorentz force}
\label{AppendixLorentzforce}
The external Lorentz force that induced by the alternating magnetic field is obviously an alternating force. However, with the considering of the huge difference in time scales between magnetic field (about $100\sim400$kHz) and flow field ($\upsilon_f/L^2$ is far more less than $1s^{-1}$). Furthermore, when time step is smaller enough than the time scale of flow field but larger enough than that of magnetic field, this alternating Lorentz force can be simplified to a steady model. The derivation process is shown as follows.

A homogeneous alternating magnetic field is assumed and an angle with x-axis is $\alpha$. The angular frequency is $\omega$, and the amplitude of magnetic field is $B_0 (B_0=H_0/\mu_0)$, so the alternating Lorentz force on x-axis is given by
\begin{align}
    F_{Mx}&=-\sigma_{nf}B_y^2u+\sigma_{nf}B_xB_yv  \notag \\ 
    &=-\sigma_{nf}B_{y0}^2\sin^2{\omega t}u+\sigma_{nf}B_{x0}B_{y0}\sin^2{\omega t}v  \notag  \\ 
    &=\sin^2{\omega t}(-\sigma_{nf}B_{y0}^2u+\sigma_{nf}B_{x0}B_{y0}v)
    \label{EqnExterFMx}
\end{align}

Then a time period of $[0,nT]$ is assumed in which $n$ cycles of magnetic field are included but the velocity $(u,v)$ of flow field is approximately unchanged. Then an integration on this time period is conducted as follows

\begin{align}
    \int_0^{nT}F_{Mx}dt&=\int_0^{nT}\sin^2{\omega t}(-\sigma_{nf}B_{y0}^2u+\sigma_{nf}B_{x0}B_{y0}v)dt \notag \\
    &=(-\sigma_{nf}B_{y0}^2u+\sigma_{nf}B_{x0}B_{y0}v)\int_0^{nT}\sin^2{\omega t}dt \notag \\
    &=(-\sigma_{nf}B_{y0}^2u+\sigma_{nf}B_{x0}B_{y0}v)\int_0^{nT}\frac{1}{2}(1-\cos{2\omega t})dt \notag\\
    &=(-\sigma_{nf}B_{y0}^2u+\sigma_{nf}B_{x0}B_{y0}v)\cdot\frac{1}{2}nT \notag\\
    &=\frac{1}{2}\int_0^{nT}(-\sigma_{nf}B_{y0}^2u+\sigma_{nf}B_{x0}B_{y0}v)dt
    \label{EqnExterFMxInt}
\end{align}
where $B_{x0}$ and $B_{y0}$ are the amplitude of magnetic field in x and y directions, respectively. Obviously, $B_{x0}=B_0\cdot \cos{\alpha}$, and $B_{y0}=B_0\cdot \sin{\alpha}$. Considering the integration form of momentum equation, this deviation means the oscillation Lorentz force can be substituted by a steady Lorentz force with half that amplitude. By which, it can be proved that the oscillation magnetic intensity can be converted into a same steady magnetic intensity times parameter $1/\sqrt{2}$ as 
\begin{equation}
    F_{Mx}=-\sigma_{nf}(\frac{1}{\sqrt{2}}B_0)^2\sin^2\alpha u+\sigma_{nf}(\frac{1}{\sqrt{2}}B_0)^2\sin \alpha \cos \alpha v
\end{equation}

The alternating Lorentz force on y-axis is given by
\begin{align}
    F_{My}&=-\sigma_{nf}B_x^2v+\sigma_{nf}B_xB_yu   \notag\\
    &=-\sigma_{nf}B_{y0}^2\cos^2{\omega t}v+\sigma_{nf}B_{x0}B_{y0}\sin^2{\omega t}u    \notag\\
    &=\sin^2{\omega t}(-\sigma_{nf}B_{x0}^2v+\sigma_{nf}B_{x0}B_{y0}u)
    \label{EqnExterFMy}
\end{align}
and the similar integration process can come to the same conclusion.

According the huge time scale gap between magnetic field and flow field, the model proposed above that using a steady Lorentz force model to substitute the high frequency alternating Lorentz force. Based on this method, a validation is conducted here to test the feasibility.
The validation is divided into two steps. The first step is to validate the external force exerted by the outer steady magnetic field. The second step is to import this field information in to a same physical model but with an oscillation magnetic field where the amplitude of oscillation magnetic intensity is the $\sqrt{2}$ times as that of steady magnetic field . 

\begin{figure*}[htpb]
    \centering
    \includegraphics[width=\textwidth]{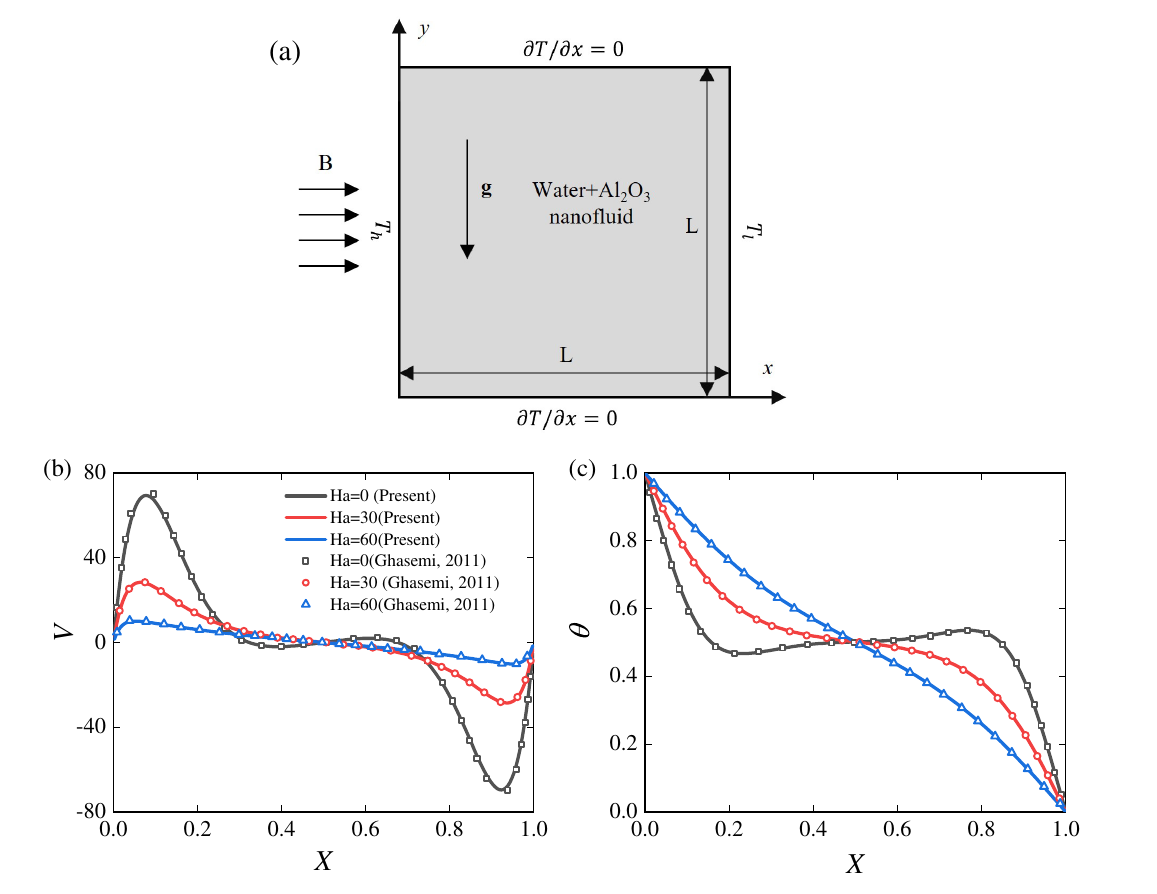}
    \caption{Validation on steady Lorentz force. (a) Schematics of natural convection with effect of magnetic field.  (b) vertical velocity on $y=0.5$, (c) temperature on $y=0.5$}
    \label{FigLorentz}
\end{figure*}

The validation on the steady external magnetic fields is conducted as the Figure \ref{FigLorentz} (a). This is a typical natural convection with the effect of uniform steady magnetic field \cite{ghasemi2011magnetic}. In the square enclosure intersperse the homogeneous nanofluid (water and Al$_2$O$_3$). The horizontal walls are adiabatic, and the constant temperature is imposed on the left wall ($T_h$) and right wall ($T_l$). All these four walls are considered as no no-slip boundary in flow field. The velocity and temperature profiles are shown as Figure \ref{FigLorentz} (b) and (c), where $Ra=10^5$, $\phi=0.03$. Results demonstrate the simulation on the different Hartman number and they match well with the Reference. \cite{ghasemi2011magnetic}.

Then the velocity and temperature information is extracted and import a same natural convection enclosure but converting the steady magnetic field into an oscillation magnetic field (from the governing equation, this change will only influence the external Lorentz force that induced by magnetic field), where the intensity of oscillation field is $\sqrt{2}$ times the intensity of steady field. After $2\cdot10^6$ cycles of magnetic oscillation, the data still keep unchanged, that means two types of magnetic field generates the same effect when the frequency of oscillation is high enough. Therefore, the steady magnetic Lorentz force model is utilized in this study.

\begin{figure*}[htpb]
    \centering
    \includegraphics[width=0.9\textwidth]{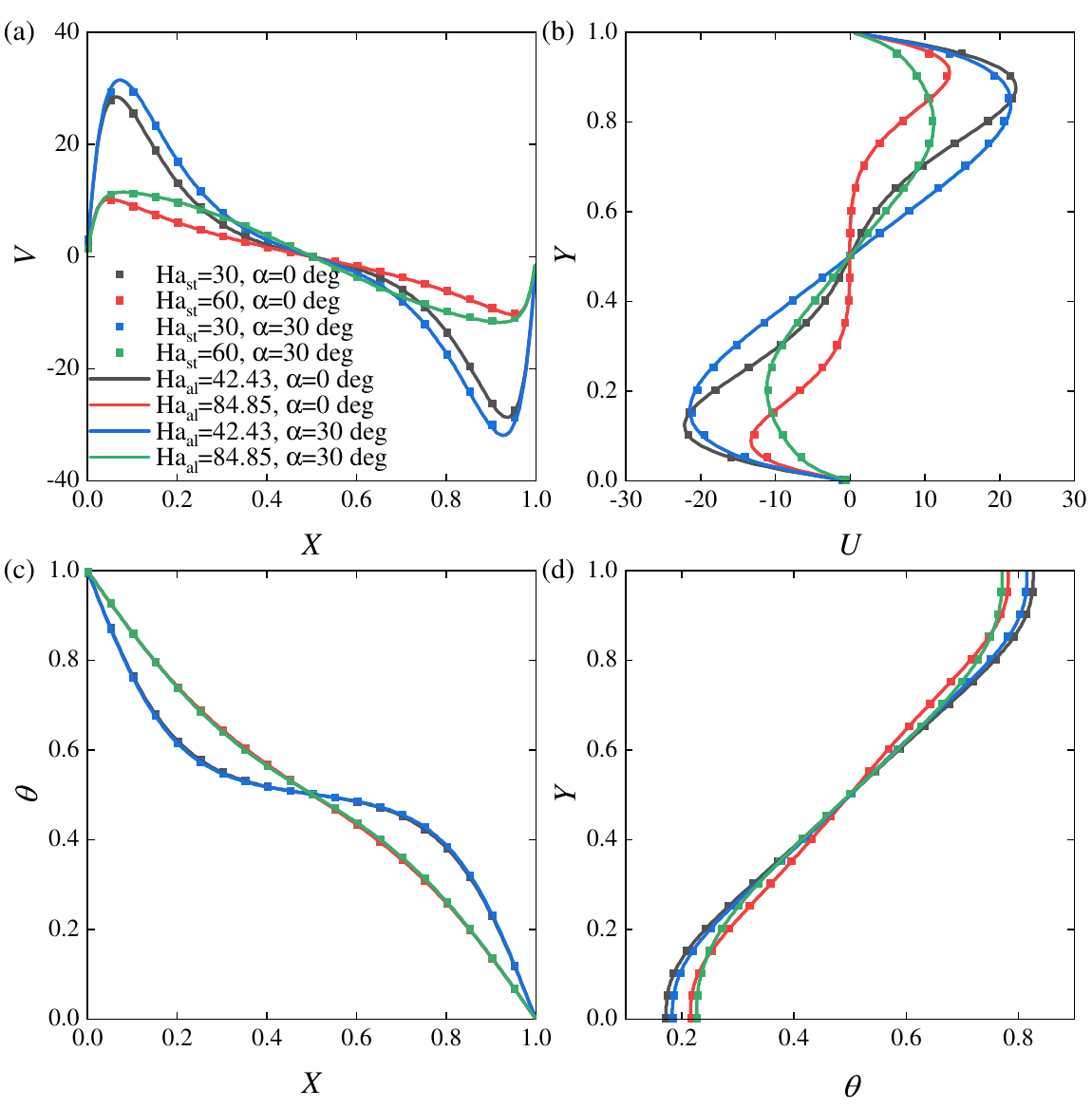}
    \caption{Validation on high frequency alternating Lorentz force. (a) vertical velocity on $y=0.5$, (b) horizontal velocity on $x=0.5$, (c) temperature on $y=0.5$ (c) temperature on $x=0.5$}
    \label{FigAlterLorentz}
\end{figure*}

 \bibliographystyle{elsarticle-num} 
 \bibliography{ref}





\end{document}